\documentclass[prb,twocolumn,aps,longbibliography]{revtex4-2}
\usepackage{graphicx,amsmath,xcolor}
\usepackage{hyperref}
\usepackage{subcaption}
\usepackage{float}
\usepackage{svg}
\usepackage{enumerate}
\usepackage{silence}
\usepackage{mathtools}
\usepackage[T1]{fontenc}
\DeclarePairedDelimiterX\innerp[2]{\langle}{\rangle}{#1\delimsize\vert\mathopen{}#2}%
\DeclarePairedDelimiterX\braket[2]{\langle}{\rangle}{#1\delimsize\vert\mathopen{}#2}%
\DeclarePairedDelimiterX\braketOP[3]{\langle}{\rangle}{#1\,\delimsize\vert\,\mathopen{}#2\,\delimsize\vert\,\mathopen{}#3}%
\DeclarePairedDelimiterX\ketbra[2]{\lvert}{\rvert}{#1\delimsize\rangle\!\delimsize\langle#2}%
\DeclarePairedDelimiterX\outerp[2]{\lvert}{\rvert}{#1\delimsize\rangle\!\delimsize\langle#2}%
\DeclarePairedDelimiterX\projector[1]{\lvert}{\rvert}{#1\delimsize\rangle\!\delimsize\langle#1}%
\newcommand{\comment}[1]{}

\begin{document}

    \title{Vision transformer based Deep Learning of Topological indicators in Majorana Nanowires}
    \author{Jacob R. Taylor}
    \author{Sankar Das Sarma}
    \affiliation{Condensed Matter Theory Center and Joint Quantum Institute, Department of Physics, University of Maryland, College Park, Maryland 20742-4111 USA}
\begin{abstract}
%Temporary Abstract
    1D superconductor-semiconductor nanowires are the leading candidates for topological quantum computation due to their ability to host non-Abelian Majorana zero modes (MZMs). However, the standard methods for identifying MZMs are often inadequate, particularly in the presence of disorder, where many properties considered to be heralds of MZMs are often generated by trivial disorder induced Andreev bound states. Recent works clearly indicate the need for developing new techniques for identifying and diagnosing MZMs. In this study, we utilize a generalized Vision Transformer-based neural network to predict, using tunnel conductance measurements, both whether a device manifests a topological MZMs phase in the presence of disorder, and also to map out the entire topological phase diagram. We show the ability of our method up to arbitrary confidence ($P>0.9998$) in classifying a device as possessing a non-trivial MZM-carrying topological phase for a wide variety of disorder parameters. We demonstrate an ability to predict from conductance measurements alternative (to the extensively used scattering-matrix-invariant topological indicator) Majorana indicators based on local density of states (LDOS). This is relevant since topology may not be uniquely defined by the scattering invariant in short disordered wires. This work serves as a significant advance offering a step towards the practical realization of Majorana-based quantum devices, enabling a deep-learning understanding of the topological properties in disordered nanowire systems. We validate our method using extensive simulated Majorana results in the presence of disorder, and suggest using this technique  for the analysis of experimental data in superconductor-semiconductor hybrid Majorana platforms.
\end{abstract}
\maketitle 

\section{Introduction}

1D superconductor-semiconductor nanowires are considered the leading platform for realizing topological quantum computation due to their ability to host Majorana zero modes (MZMs) 
\cite{lutchyn2010majorana,sau2010robustness,oreg2010helical,sau2010generic,das2023search,sarma2015majorana,lutchyn2018majorana}. These non-Abelian quasiparticles are protected by an energy gap ("topological gap") from local perturbations due to their non-local topological encoding of quantum information \cite{kitaev2001unpaired}. This topological protection allows the execution of error-free gate operations, such as braiding, enabling fault-tolerant quantum computing. By harnessing these properties, 1D nanowires offer a promising pathway to developing robust quantum computing systems capable of overcoming many of the serious error-correction challenges faced by traditional quantum computing approaches. Several experiments have reported the observation of zero-bias conductance peaks (ZBCPs) that are compatible with the presence of Majorana zero modes in 1D superconductor-semiconductor nanowires \cite{das2012zero,deng2012anomalous,mourik2012signatures,churchill2013superconductor,finck2013anomalous,nichele2017scaling,zhang2021large,yu2021non,song2022large,aghaee2023inas}. The severe problem with these devices, however, is that the existence of disorder arising mainly from unintentional random charged impurities \cite{ahn2021estimating} can destroy the non-trivial topological properties particularly for strong disorder compared with the superconducting proximity gap in the nanowire. More confounding, disorder can mimic behaviors such as the appearance of ZBCP, which are typically considered signatures of Majorana zero modes \cite{das2023search}. In a clean, disorder-free system, a zero-bias peak in the conductance spectrum is expected to arise from the presence of a Majorana mode at the end of the nanowire, which should also be quantized at low temperatures \cite{sengupta2001midgap}. However, in disordered systems, similar peaks can be produced by trivial bound states or Andreev bound states, which are not topologically protected and do not correspond to true MZMs \cite{pan2020physical,liu2012zero,das2021disorder}. These disorder-induced bound trivial states may accidentally localize near the ends of the wire, mimicking the spatial localization expected for Majorana modes. Such accidentally end-localized trivial bound states may also produce misleading topological signatures ("false positives") by satisfying the standard scattering matrix related invariant indicators for MZMs, which are often used as Majorana diagnostics in theories and simulations. Consequently, these false positives can mislead experimental interpretations, suggesting the presence of MZMs where none exist \cite{pan2020physical,liu2012zero}. This is an open key problem in the subject. The challenge is that topology itself cannot be directly measured through transport spectroscopy, and has to be inferred from its predicted theoretical signatures, which are often also produced by disorder.

Several experiments have reported the observation of zero-bias conductance peaks (ZBCPs) that are compatible with the presence of Majorana zero modes (MZMs) in 1D superconductor-semiconductor nanowires \cite{das2012zero,deng2012anomalous,mourik2012signatures,churchill2013superconductor,finck2013anomalous,nichele2017scaling,zhang2021large,yu2021non,song2022large,aghaee2023inas}. It is now generally accepted that essentially all of these experiments observed the 'fake' MZMs mimicked by disorder induced Andreev bound states, particularly since no special efforts were made to obtain very clean systems in these experiments. One recent experiment may have partially remedied this disorder problem using extremely clean starting materials and through validation with the topological gap protocol \cite{aghaee2023inas}, a procedure that provides strong evidence of MZMs detection through an automated protocol \cite{pikulin2021protocol}. In addition, the typical nanowire superconducting gaps are rather small, further exacerbating disorder problems and the associated confusion caused by disorder-induced trivial states in the interpretation.  The problem of disorder and the associated difficulties in the identification of clear signatures for MZMs have strongly hindered progress in the subject as has recently been emphasized \cite{das2023search}. 

There has been significant work dedicated to correcting and understanding the effects of disorder in nanowire Majorana platforms \cite{taylor2024machine,thamm2023machine}. In particular, efforts have focused on characterizing how disorder impacts the utility of 1D nanowire devices for topological computation. The primary indicator for topology currently used is the topological visibility or scattering invariant, which determines whether the wire contains a topologically non-trivial state \cite{akhmerov2011quantized,kitaev2001unpaired}. The scattering invariant is crucial because it serves as a unique criterion for identifying Majorana modes in very long low-disorder wires \cite{das2016infer,das2023spectral}, and is also a necessary (though not sufficient) criterion for detecting MZMs in short disordered wires. Unfortunately, the scattering invariant cannot be directly measured in experiments and is instead estimated using device simulations and heuristic methods, which have limitations \cite{pikulin2021protocol,aghaee2023inas}. The problem thus stills exists in effectively characterizing the topology of the experimental devices and addressing this gap by finding a reliable method to determine the topological invariant remains critical for the field \cite{das2023spectral,das2023density,pan2024disordered}.

As described above and as emphasized in many recent publications \cite{das2023search,peeters2024effect,sau2010generic,das2012zero,zhang2021large,yu2021non,pikulin2021protocol,ahn2021estimating,das2016infer,das2023spectral,cheng2024machine,pan2024disordered,groth2014kwant}, disorder is the most important roadblock in further progress in topological Majorana physics, but an equally important complementary problem receiving much less attention is figuring out the appropriate signatures and diagnosis of topolgical MZMs.  The original idea of just focusing on the appearance of experimental zero bias tunnel conductance peaks in a finite magnetic field as the signature of MZMs has failed spectacularly as it is now obvious that zero bias tunnel conductance peaks are generically produced by disorder \cite{das2023search}.  Most, if not all, of the experimentally reported zero bias peaks are likely disorder-induced trivial subgap bound states.  It is now clear that zero bias conductance peaks may be necessary, but by no means sufficient, in providing signatures of MZMs, and at the minimum, some bulk signatures are necessary for topology.  Possible such bulk MZM signatures are the MZM splitting oscillations \cite{das2012splitting},  zero bias peak correlations from both ends \cite{lai2019presence}, and the manifestation of the closing/opening of a bulk gap associated with the topological quantum phase transition \cite{pan2021three,vuik2019reproducing,lai2019presence}.  Only the most recent experiment from Microsoft goes beyond just the simple zero bias conductance peak measurements, presenting evidence for end-to-end zero bias correlations as well as for the closing/opening of a bulk gap \cite{aghaee2023inas}.   The serious question still remains open whether such necessary conditions (e.g. ZBCP, end-to-end correlated ZBCPs, gap closure and reopening features in the nonlocal conductance, etc.) are sufficient to claim the existence of non-Abelian MZMs in the sample which can be used to carry out fault-tolerant topological quantum computation \cite{das2023spectral,das2023density}.  In particular, some other necessary conditions such as the precise quantization of the ZBCP at the value of $2e^2/h$ and the Majorana splitting oscillations are typically absent from the experimental data.

What is needed is a definitive demonstration of signatures of nontrivial topology in the experimental data, but this is hard to come by until specific braiding measurements are done demonstrating non-Abelian anyonic properties. Such braiding measurements are far in the future necessitating some new topological characterization of the readily-available tunnel conductance measurements.  The best current option, which has been used as the main diagnosis in the recent Microsoft experiment \cite{aghaee2023inas} through the topological gap protocol, is to carry out numerical simulations of the experiment using as realistic a model of the nanowire system as possible.  By construction, one knows whether the system is topological or not in a simulation, and the hope is that such a simulation can then tell us what is happening in the experimental system.  The problem however, is that, as emphasized in several recent publications \cite{das2023spectral,pan2024disordered,das2023density}, there is no unique topological signature for short disordered nanowires since topology is by definition a nonlocal property manifest only at very large length scales.  In particular, disorder and the observed small superconducting gap in the nanowires makes the coherence length rather long, comparable to or perhaps larger than the wire length (or the localization length whichever is shorter), and in such situations the concept of topology becomes problematic and ambiguous for the finite system.  The most obvious (and the most extensively used) signature for topology is the scattering matrix invariant, which defines topology just by its sign: A negative (positive) value defines the system to be topological (trivial).  Although in very long wires, simulations indeed show that this quantity goes from +1 in the trivial regime at low magnetic field to -1 in the topological regime at high magnetic field with the sign change happening abruptly and precisely at the topological quantum phase transition point, this nice dichotomy does not apply in finite disordered systems where the scattering matrix invariant is quite complicated, manifesting continuous behavior with sometimes multiple sign changes! \cite{pan2024disordered}  Therefore, one needs better topological signatures even for the simulations, let alone in the experiments, and using just the topological visibility (TV) arising from the scattering matrix invariant is not enough \cite{akhmerov2011quantized,das2023spectral}.

In the current work, we propose combining the latest developments in the machine learning of image processing ("vision transformer") \cite{dosovitskiy2020image} along with multiple topological identifiers for MZMs \cite{das2023spectral,pan2024disordered} in order to determine whether specific conductance results in a nanowire represent underlying topological non-Abelian Majorana zero modes. The idea here is treating the nanowire conductance results as a set of images where computer vision machine learning techniques can directly provide the underlying topological information somewhat reminiscent of the way image processing leads to facial recognition in the real world. Neural network models are capable of analyzing data in a way that captures complex relationships, making them particularly effective in handling systems where traditional methods of analysis fail. Recently, there has been a growing interest in studying disordered physical systems—where traditional approaches are difficult or impossible to apply—through the use of machine learning. A scheme proposed in \cite{taylor2024neural} outlines a method for determining disorder through easily performed device measurements and demonstrates this by identifying the full disordered Hubbard model parameters as appropriate for semiconductor spin qubits. In another study, this method was applied to show that it is, in principle, possible (though numerically challenging) to determine significant disorder properties of a 1D nanowire device using only conductance measurements \cite{taylor2024machine}. Current approaches to understanding the effects of disorder on the topology of these devices often rely on heuristic methods, which involve visually analyzing multiple conductance plots. For instance, in the topological gap protocol\cite{aghaee2023inas}, when made quantitative, a series of visual processing algorithms are implemented using arbitrary thresholds. These thresholds can be adjusted, potentially leading to devices with non-topological properties being incorrectly identified as topological \cite{aghaee2023inas}. Since these thresholds are set manually, there is considerable skepticism about whether any device declared topologically non-trivial is genuinely topologically non-trivial \cite{das2024comment}. Although some simulations have tested the accuracy of the topological gap protocol on simulated devices \cite{aghaee2023inas}, doubts persist, particularly because the magnitude of the disorder itself remains unknown and because the typical extracted topological gaps are small ($\approx 20 \mu$eV).

Since, in practice, all heuristics on the topological properties of 1D nanowires currently involve some form of visual observation on conductance data and setting arbitrary thresholds—whether quantitatively or unknowingly qualitatively—on whether a device is topological or not, it makes sense to utilize a neural network tool that is particularly well-suited for this task. Vision transformers, with their ability to capture and process intricate patterns within measurement result landscapes, offer a well-defined and non-arbitrary method for diagnosing topological invariants from the visual analysis of large amounts of conductance data. This is achieved by leveraging the transformer architecture's strength in handling spatial correlations and locality within the data.

Here, we seek to combine advances in computer vision, particularly the use of vision transformers \cite{dosovitskiy2020image}, to analyze conductance measurements, which are routinely performed on MZMs devices, to directly determine the topological invariant. We approach this problem in a continuous manner, enabling the prediction of the actual value of the scattering invariant rather than just its sign. This continuous determination offers advantages over discrete classification models by allowing for the introduction of a topological cutoff.  By tuning this cutoff, our method provides a way to classify a device as topologically non-trivial with an arbitrarily low probability of false positives. This means that the chance of incorrectly declaring a device as topologically non-trivial can be made arbitrarily small. { While in classification the softmax layer output can also be used to set probabilities it would lose physically insightful information of the indicator.} Furthermore, determining the topological invariant continuously is crucial because the sign of the scattering invariant is only a perfect criterion for an infinitely long wire. In finite wires, where disorder suppresses topological properties, both the sign and magnitude of the topological invariant should be determined to prevent false positives.

In prior machine learning (ML) work aimed at determining whether a device possesses MZMs, only devices with fixed high or low levels of disorder were considered \cite{cheng2024machine}. These earlier studies therefore mostly focused on the relatively easy findings of 'topological' (for low disorder) and 'trivial' (for high disorder), leaving out the difficult intermediate disorder regime which may be of experimental relevance. We extend this analysis by evaluating the topological invariant across a broad range of disorder regimes by using the vision transformer technique. Specifically, we examine three regimes: the {extremes regime} (where disorder is either very high or very low, similar to previous studies), the full regime (a continuum from low to high disorder), and the moderate disorder regime (where disorder is fixed at a moderate or intermediate strength, which is the most challenging to predict but closely aligns with current experimental progress). Additionally, we explore a wide variety of unknown parameters, including the disorder correlation length and the (often unknown) spin-orbit coupling constant $\alpha$, both of which are difficult to measure experimentally. This approach assesses the robustness of our ML scheme in determining the topological invariant across these varied disorder regimes.

Following this, we will significantly extend upon prior work \cite{cheng2024machine} by demonstrating our methods' ability to generate the entire topological phase diagram for different experimental parameters, allowing one to determine the location of the topological phase precisely and potentially giving experimentalists a new and currently unprecedented ability to understand the topological properties of their devices through their applying our technique to the experimental data analysis.

While it may seem as if this vision transformer machine learning technique by itself can solve the topology question in experimental Majorana nanowire devices, this is not entirely true. Recent work \cite{pan2024disordered} suggests that being able to determine just the scattering invariant may be insufficient to determine whether a 1D nanowire can be used for fusion and braiding and since these are the basis of topological quantum computation, it too. In short disordered wires, the scattering invariant may have to be complemented by other operational topological indicators.

These alternative operational indicators proposed in \cite{pan2024disordered} provide not only an assessment of whether a topological state exists but also whether such a state is usable for topological computation. These operational indicators help prevent false positive declarations by ascertaining whether the Majoranas are localized at the ends without internal domain walls that could potentially break the wire into a series of isolated quantum dots. While these indicators seem capable of accurately assessing whether a device is topological when combined with the scattering invariant, there is currently no practical direct method for measuring them experimentally since they involve a knowledge of the local density of states along the wire. With this in mind, we as a first in its class result provide a novel method to determine these additional false positive-preventing indicators directly from conductance measurements. Our technique using only the conductance data should be usable to analyze future experiments to directly conclude about their topological properties. 

The paper will be organized as follows:

\begin{enumerate}[I]
    \item \textbf{Introduction}
    
    \item \textbf{Model}: Provides details about the model used. 
    \item \textbf{Method}
        \begin{enumerate}[A]
        {\item \textbf{Majorana Indicators}: Provides details about the Majorana indicators which will be predicted.}
        \item \textbf{Training Data}: Explains training data generation and setting up the disorder for the machine learning problem.
        \item \textbf{Neural Network}:  Discusses the deep learning neural network architecture.
        \item \textbf{Disorder Regimes}: Discusses the different parameter regimes considered. 
    \end{enumerate}
    \item \textbf{Results}         
        \begin{enumerate}[A]
            \item \textbf{Aggregated Topological Invariant}: Presents demonstration of the ability of the scheme to predict the scattering invariant of a device. 
            \item \textbf{Continuous Topological Invariant}: Presents demonstration of the ability of the scheme to predict the entire phase diagram of the scattering invariant for a device.
            \item \textbf{Continuous Alternative Indicators}: Presents demonstration of the ability of the scheme to predict the entire phase diagram of alternative Majorana indicators for a device.
        \end{enumerate}
    \item \textbf{{Summary and }Conclusion}
\end{enumerate}
%--------

\section{Model}
We model the 1D semiconductor Majorana nanowire using a Bogoliubov-de Gennes  Hamiltonian \cite{lutchyn2010majorana}:
\begin{multline}\label{eq:1}
H=\left(-\frac{\hbar^2}{2m^*}\partial_x^2-i\alpha\partial_x\sigma_y-\mu+V_{\text{dis}}(x)\right)\tau_z+\\\frac{1}{2}g\mu_BB\left(\sigma_x\cos\theta+\sigma_y\sin\theta\right)+\Sigma(\omega)\end{multline} where $\Sigma(\omega)=-\gamma \frac{\omega+\Delta_0\tau_x}{\sqrt{\Delta_0^2-\omega^2}}$ is the self-energy generated by integrating out the superconductor proximity coupled to the semiconductor~\cite{sau2010robustness}. $\sigma_{x,y,z}$ and $\tau_{x,y,z}$  are the Pauli matrices for the spin degree of freedom, and the superconducting particle-hole degree of freedom respectively. The above Hamiltonian is written in a basis where the Bogoliubov quasiparticles of the superconductor are described by a Nambu spinor with a structure $\psi(x)=(u_{\uparrow}(x), u_{\downarrow}(x),-v_{\downarrow}(x),v_{\uparrow}(x))^T$, where $\uparrow,\downarrow$ refer to spin and $u,v$ refer to particle-hole components of the quasiparticle. The frequency $\omega$ in the self-energy is equivalent to the energy (in units where $\hbar=1$) of the Bogoliubov quasiparticle in consideration. We mention that Eq. \ref{eq:1} represents the standard Majorana nanowite model used extensively in the literature.

%-Add bit splitting hamiltonian into parts, like haining always does. 
%...

%CHECK CITATIONS ON BELOW
KWANT solves the transport problem of the Majorana nanowire by computing the scattering matrix of the system in the particle-hole space between the two leads to determine the conductance matrix via the well-known Blonder-Tinkham-Klapwijk relations \cite{groth2014kwant}. %In practice KWANT makes use of a wave function matching ... %It also uses the reflection matrix to compute topological visibility\cite{akhmerov2011quantized,pan2021three}. 
Despite the complexity of the actual semiconductor device, KWANT's model Hamiltonian captures the experimental quantitative transport features when parameters such as effective mass, spin-orbit coupling, and superconducting pairing potential are treated as fitting parameters \cite{groth2014kwant}. Specific parameter values and conditions (e.g., effective mass \(m^*=0.03 m_e\), superconducting pairing potential \(\gamma=0.15\,meV\), g-factor \(g=25\), superconducting gap \(\Delta_0=0.12\,meV\), temperature \(T=50mK\), and device length \(L=3\,\mu m\)) are chosen to fit non-superconducting transport characterization and superconducting phase transport properties \cite{pan2021three,woods2021charge,das2023spectral}. The spin-orbit coupling \(\alpha\approx 8.0\,meV*nm\) is similarly fit but is less apparent experimentally, for this reason within our results we will consider this a random parameter which will be varied within our training data realizations to ascertain its value using machine learning.  Note that our choice of these parameters corresponding to the currently used InAs-Al nanowires is just a matter of convenience, and obviously changing thee parameters  to other values corresponding to other materials does not change anything-- all one needs to do is to change the parameters and reobtain the training data. The finite temperature effect is modeled by convolving zero-temperature conductance with the Fermi function derivative. The Hamiltonian is discretized with a lattice scale of 10 nm which should be sufficient and is based on prior work \cite{das2023spectral} aimed at understanding recent experiments \cite{aghaee2023inas}. Again, this choice is independent of our machine learning scheme, and can be whatever is appropriate for simulating the system under consideration.

The most difficult and completely unknown fitting parameter in our model is that of the spacially varying random disorder potential $V_{\text{dis}}(x)$. The disorder potential generated by impurities in the device is incorporated into the Hamiltonian \(H\) through the function \(V_{\text{dis}}(x)\). This spatial disorder potential varies randomly with position \(x\) along the wire, following a Gaussian distribution with some amplitude and correlation length which is not actually known. While, experimentally both the disorder magnitude and correlation length can be roughly estimated from the density of charge defects determined from self-consistent electrostatics calculations \cite{winkler2019unified}, such methods are limited in any quantitative accuracy as often the actual impurity locations are unknown.  In our work, we use random $V_{dis}(x)$ to generate our training and test data for machine learning. We also include barrier potentials $V_{\text{Barrier}}^L$ and $V_{\text{Barrier}}^R$ as $\delta$ function contributions at the ends of the wire to $V_{\text{dis}}(x)$, we treat these barriers as fixed but tunable parameters. These barrier voltages will be set such that $V_{\text{Barrier}}^L=V_{\text{Barrier}}^R=15$mV. Of course, it is easy to vary them if necessary at the cost of generating additional training and test data. To recover information about a device we will vary experimentally tunable parameters within the device, in particular the chemical potential $\mu$ tunable by plunger gates and the external magnetic field $B$ (which tunes the Zeeman energy). We will vary $B$ between [0.0T,0.8T] with 20 points and $\mu$ between [0.2meV,0.4meV] with 5 point. These ranges are motivated by current experiments. For each different configuration of tunable experimental parameters we will perform differential conductance measurement simulations based on Eq. \ref{eq:1} to gather information about the properties of each 1D nanowire device.  The differential conductances are defined as $G_{\alpha \beta}= dI_\alpha/dV_\beta$ for $\alpha =\beta$ and $G_{\alpha \beta}= -dI_\alpha/dV_\beta$ for $\alpha \neq \beta$. Local conductances $G_{LL}$ and $G_{RR}$ are measured from their respective leads through the junction (to the superconductor), while non-local conductances $G_{RL}$ and $G_{LR}$ are measured from the right lead to the left lead or the reverse respectively. We make use of all 4 of these conductance measurements within the machine learning scheme. We allude to our calculated conductance results as experimental 'measurements' in the spirit of our machine learning goals since these calculated results ("measurements") serve only as the data used for training the neural networks (and then to test its predictive abilities for test 'measurements' outside the training data)  These KWANT-based simulations aof Eq. \ref{eq:1} are standard for studying nanowire MZMs and have been used extensively.  The only point to make is that for us these simulations provide both the training and the testing data to carry out our machine learning and check its validity.

The generated differential conductances from KWANT are computed initially for fixed $\omega$ at zero temperature, however to allow finite temperature a $\omega$ convolution is performed with the Fermi distribution f(E) as follows \cite{setiawan2017electron}:
$$G(V_{Bias})=-\int_{-\infty}^\infty d\omega G_0(\omega)\frac{df(\omega-V_{Bias})}{d\omega}$$

where $G_0$ is the zero temperature conductance. This finite temperature convolution requires aggregating conductance measurements for many $\omega$ or $V_{Bias}$ values. Within our simulations we consider $V_{\text{Bias}}$ values between $\pm 0.05 meV$ with 81-151 points. This implies we are required when accurately simulating the physics of such devices to generate a large amount of conductance measurements, just from the  $V_{\text{Bias}}$ requirements. We found no meaningful difference in the neural networks fidelity by changing the number of $V_\text{Bias}$ points within this range. The finite temperature conductances are the only conductances fed into the neural network. A huge set of such conductance data are generated numerically in order to carry out the machine learning procedure using the vision transformer.
{\section{Method}}

{\subsection{Localized Majorana Indicators}}

To evaluate a 1D nanowire's suitability for fusion and braiding, metrics that are not experimentally accessible but theoretically indicative of utility are commonly used. The most important and frequently utilized metric is the scattering invariant, or topological visibility \cite{akhmerov2011quantized,das2016infer}. This invariant is defined as the sign of the determinant of the scattering matrix:

$$T_V=\det(\begin{bmatrix}r_L^{e\rightarrow e} & r_L^{e\rightarrow h} \\ r_L^{h\rightarrow e} & r_L^{h\rightarrow h}\end{bmatrix})$$
$$Q=\text{sgn} \left(T_V\right)$$

Where $r_{L}^{i\rightarrow j}$ are the reflection coefficients in the scattering matrix from $i \in {h,e}$ to $j \in {h,e}$, where {h,e} denote hole quasi-particles and electrons, respectively. Similar to conductance, the reflection matrix can be computed directly within KWANT \cite{akhmerov2011quantized,pan2021three}. A negative value of the metric $Q$ indicates that the wire is in the topologically non-trivial regime, while a positive value indicates the wire is topologically trivial. The scattering invariant works by allowing one to probe the transition from normal reflection to Andreev reflection and thus the transition to a topologically non-trivial state \cite{akhmerov2011quantized,das2016infer}. We also define a continuous version of this metric, known as topological visibility $T_V$. This is necessary because, while a negative value in the scattering invariant is sufficient to declare a wire topologically non-trivial in the infinite length limit, this may not hold true for finite-length wires where the sign alone may not capture whether the topological properties occur \cite{pan2024disordered}. Thus, the actual magnitude of the topological visibility becomes important. Topological visibility is the metric most used in theoretical work on Majorana nanowire systems and is completely equivalent to the Pfaffian topological indicator of Kitaev \cite{kitaev2001unpaired}.

The topological visibility is a metric for determining whether non-local Majorana pairs exist; however, although it can confirm the presence of a MZMs, it does not ensure the utility of those Majoranas for practical applications. For non-Abelian anyon-based topological quantum computing, the MZMs need isolated zero-energy states localized at the ends of the wire for braiding and fusion \cite{das2023spectral,pan2024disordered}. This requirement is especially critical for experimentally relevant short, disordered wires, where topological properties may be suppressed due to disorder. In particular, one serious problem pointed out in recent work \cite{das2023spectral,pan2024disordered} is the possible disorder-induced presence of multiple almost zero-energy Andreev bound states in the bulk of the wire in addition to the end Majorana modes.  Such Andreev states in the bulk of the wire cannot be detected simply by the topological visibility. Recent research has therefore explored alternative metrics for assessing the presence of localized Majoranas at the ends of the wire using the local density of states (LDOS), defined as $\rho =\sum_n\langle\psi_n|r\rangle\langle r|\psi_n\rangle\delta(E-E_n)$ \cite{das2023density,pan2024disordered}. The LDOS provides a method to understand the properties of the spatial distribution of the subgap states and can indicate whether Majoranas are localized at the wire's ends. Practically, the LDOS is not directly accessible experimentally, similar to the scattering invariant (though it could, in principle, be measured by scanning tunneling spectroscopy) \cite{pan2024disordered}. It may, however, be accessible through machine learning by utilizing the information contained in conductance measurements. We will use recently proposed alternative indicators motivated by the fact that well-defined localized MZMs should have an LDOS that vanishes in the bulk and is large at the ends \cite{pan2024disordered}. Specifically, we will consider two alternative indicators, $I_{2}^{(4)}$ and $I_{1}^{(4)}$ \cite{pan2024disordered}. To use these indicators, we need to define edge and bulk sections of the wire. Edge sites are those within $\epsilon/2$ of the wire's edge, where $\epsilon = 70 \text{ nm}$ is approximately the coherence length obtained from fitting the disorder-free LDOS. This value does not need to be exact, as tunable thresholds will be used to determine if the indicator is satisfied. This is particularly relevant since the number should vary with both $\mu$ and $B$, so we choose a representative value within this range. 

The $I_{2}^{(4)}$ indicator which is the ratio of maximal LDOS in the bulk over the end
LDOS, is defined as: 

{\small $$I_{2}^{(4)}=\frac{\underset{x_{i}\in\mathrm{Bulk}}{\max}\rho(\omega=0,x_{i})}{\frac{1}{2}\biggl[\underset{x_{i}\in\left[0,\frac{\xi}{2}\right]}{\max}\rho(\omega=0,x_{i})+\underset{x_{i}\in\left[L-\frac{\xi}{2},L\right]}{\max}\rho(\omega=0,x_{i})\biggr]}$$}

where $L$ is the length of the wire. This indicator functions by comparing the peak amplitude of the LDOS in the bulk of the wire to the average of the peaks at the two ends of the wire. For well-localized MZMs, this ratio should not exceed 1, and thus operationally, we will consider $I_2^{(4)} < 1$ to be a pass or true positive. The $I_{1}^{(4)}$ indicator is defined as:
{\small
$$I_{1}^{(4)}=\frac{\rho\bigl(\omega=0,x_{i}=\frac{L}{2}\bigr)}{\frac{1}{2}\biggl[
\underset{{x_{i}\in\left[0,\frac{\epsilon}{2}\right]}}{\max}\rho(\omega=0,x_{i})+\underset{x_{i}\in\left[L-\frac{\epsilon}{2},L\right]}{\max}\rho(\omega=0,x_{i})\biggr]}.$$}

This indicator works by determining whether the LDOS at the midpoint of the wire is less than the LDOS peaks at the ends by a certain threshold. For a well-localized MZM, this value should be quite small, so operationally, we will consider $I_1^{(4)} < 0.1$ to be a pass or true positive. We use these thresholds, which are determined by observing the ideal case with no disorder; however, it should be noted that setting these cutoffs does involve an element of arbitrariness. We will also shift and rescale these indicators based on the cutoff values to be between 1 and -1, such that a pass yields a negative value and a fail yields a positive value. The boundaries are achieved by truncating the indicators, as after a certain point it is not important how much an indicator fails. These truncated and rescaled indicators are defined as:

$$I_2=\min \left(I_{2}^{(4)}-1, 1\right)$$

$$I_1=\min \left(\frac{I_{1}^{(4)}}{0.1}-1, 1\right)$$

\subsection{Training Data Generation}

The generation of training data requires simulating many different realizations of 1D nanowire devices with different parameters. Each wire realization requires an independently and randomly determined $\alpha$ and a spatially dependent disorder potential $V_{\text{dis}}(x)$. Each $V_{\text{dis}}(x)$ also requires a randomly sampled disorder magnitude and correlation length, with the range of these parameters depending on the disorder regime being considered. Every device thus has a "disorder vector" comprising all the site's disorder potentials along with other relevant parameters: $\vec{D}=[\vec{V}_\text{{dis}}, \alpha, L_c, \sigma_{\text{dis}}]$ where $L_c$ is the disorder correlation length and  $\sigma_{\text{dis}}$ is the disorder magnitude. In this work, we do not attempt to determine any of these parameters and instead treat them as fully unknown. Additionally, we create a matrix listing the values of experimental parameters, such as the chemical potential $\mu$ and magnetic field $B$, to be tuned to gather information about the device.

\begin{equation}
K^j=[B^j, \mu^j, V_{Bias}^j]
\end{equation}

For each 'measurement' configuration (or row $K^j$), the conductance measurements (i.e. the calculated conductance results) are used to form an input matrix $X$. This $X$ matrix, formed from experimental measurements, will be utilized by the neural network to determine the indicators (which could be anything we desire including the scattering invariant and additional independent indicators). This $X$ matrix is shown as follows:

\begin{equation}
X^j=[G_{LL}^j,G_{LR}^j,G_{RL}^j,G_{RR}^j]
\end{equation}

Along with the conductance measurements, KWANT also produces our desired output vector of the neural network, denoted as the vector $Y$. The vector $Y$ comprises a list of all the indicator values for each relevant measurement configuration, specifically those where $V_{\text{Bias}}=0$. It should be noted that we train a new neural network for each of the indicators, rather than attempting to learn all of them at once, even though we generate the training data for all indicators simultaneously. The vector $Y$ can be written as follows:

{
\begin{equation}
Y=
\begin{cases}\vec{T_V} \\  \vec{I_1}^{(4)}\\
 \vec{I_2}^{(4)}
 \end{cases}
\end{equation}
}
%This $Y$ vector is further processed in the cases where ...
It should also be noted that we keep the full matrix $\textbf{K}$ constant for all different 1D nanowire devices. Putting these components together, we have our training data generator function, which for device $i$ is written as follows:

$$f_\text{Gen} (\vec{D}_i,\textbf{K})=[\textbf{X}_i,\textbf{Y}_i]$$

With this generator function, we are able to produce arbitrary amounts of training data, each with different $D_i$, to allow for the training of our machine learning model. The goal of the neural network is thus as follows: given $\textbf{X}_i$ and $\textbf{K}$, provide a prediction for $\vec{Y}_i$:

$$f_\text{NN}(\textbf{X}_i,\textbf{K})=\vec{Y}_i$$

In other words, our machine learning model is trained to predict Majorana indicators solely from a sequence of conductance measurements. The remarkable aspect of the technique is that there is no need for the model to estimate the unknown random parameters-- they remain unknown and hidden in our machine learning just as they are in the actual experimental measurements.  The only item that matters is the conductance matrix, which is precisely what experiments measure \cite{aghaee2023inas}.  This makes this technique readily usable in Majorana experiments since nothing other than what is measured experimentally is necessary to determine the underlying topology.

\subsection{Neural Network}
\begin{figure}[]
     \centering
     \begin{subfigure}[b]{0.99\columnwidth}
         \centering
         \includegraphics[width=\textwidth]{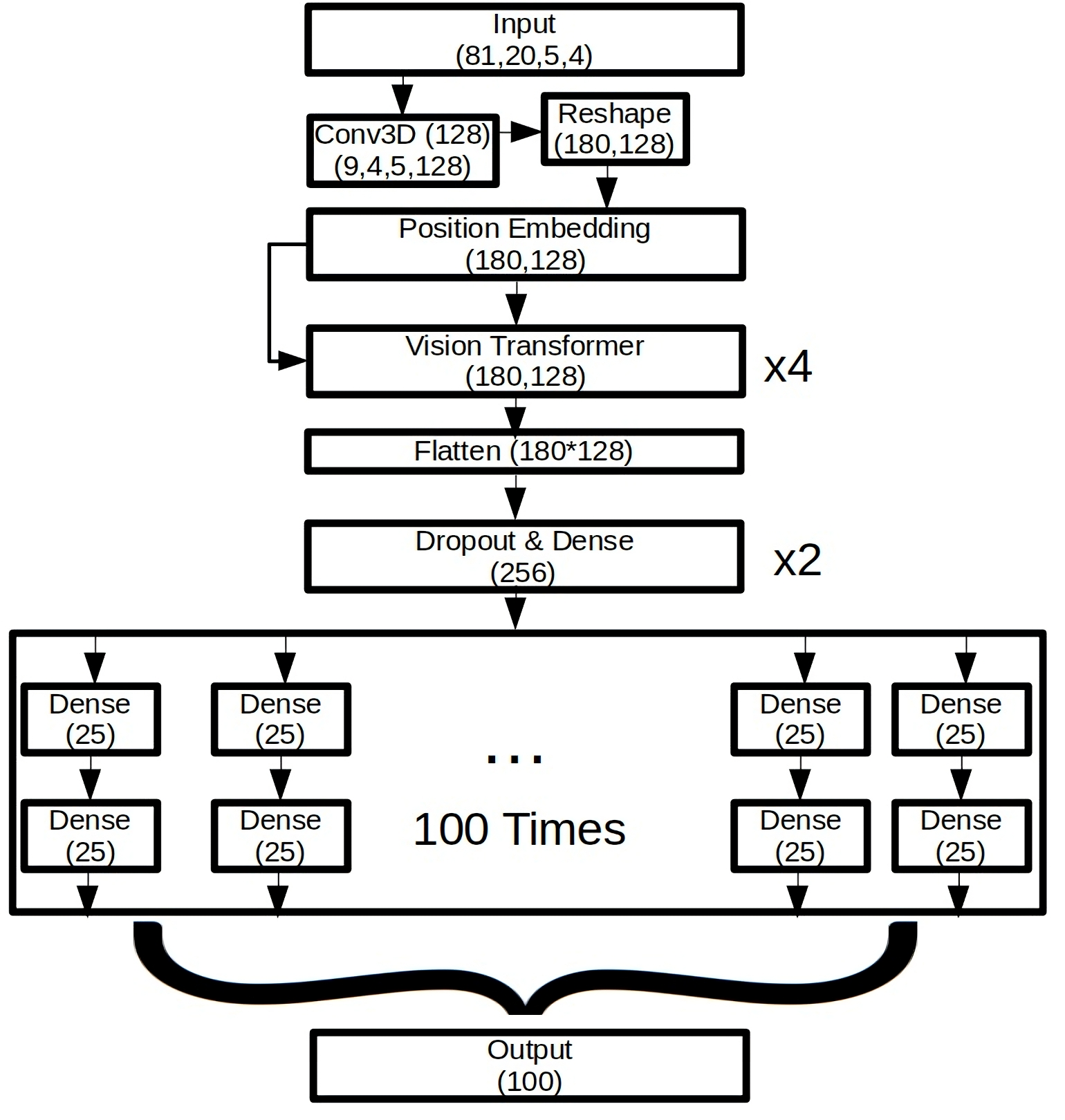}
         \caption{}
     \end{subfigure}
     \begin{subfigure}[b]{0.7\columnwidth}
         \centering
         \includegraphics[width=\textwidth]{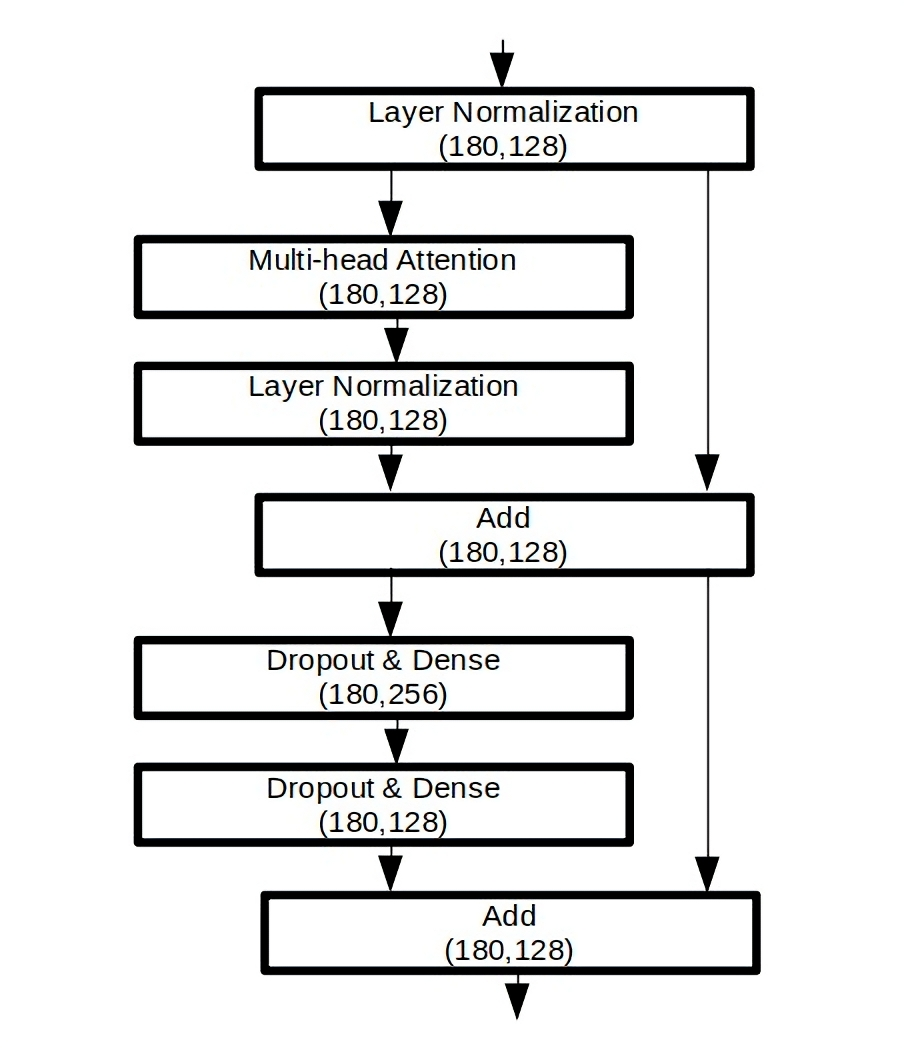}
         \caption{}
     \end{subfigure}
     \caption{(a) Diagram of the neural network used to process conductance measurements into predictions of topological/Majorana indicators. The neural network used is similar to standard vision transformer-based models, except with a many-path tree ending, allowing each point within the measurement grid (i.e., each experimental setup) to undergo some minor independent processing. Further details about the neural network's architecture can be found in the main text. (b) Diagram of the vision transformer used within (a). The vision transformer works by first performing layer normalization on the data input. It then has two additional paths: a skip path, which is used to prevent the vanishing gradient problem, and the multi-head attention path (followed by layer normalization). These two paths are combined additively. After this, another two paths form: the first where the MLP is used with dropout, consisting of two dense layers with 256 and 128 nodes, and the other another skip layer. These two paths are then also combined additively. }
     \label{fig:NNDiagram}
\end{figure}

The neural network we use is based on the Vision Transformer (ViT) model introduced in \cite{dosovitskiy2020image}. The schematics for the ViT are shown in Fig. \ref{fig:NNDiagram}. Multi-head attention, or transformers, have proven exceptionally effective in a range of applications, including natural language processing (NLP), computer vision, time series forecasting, and even reinforcement learning. Their strength in solving computer vision problems, in particular, makes them well-suited for handling problems where the input consists of arrays of measurement results with inherent locality in the data. In this context, locality pertains to the tunable experimental parameters, such as the magnetic field $B$, chemical potential $\mu$, or bias voltage, where local regions within the array correspond to similar values of these parameters.

In our case, the input is a series of conductance measurements for different experimentally tunable parameters. These measurements form a sequence of "images" where each conductance measurement type ($G_{LL},G_{LR},G_{RL},G_{RR}$) can be thought of as a color or channel. However, an immediate complication arises because, as mentioned, we are varying 3 parameters, not two, as is typical in a ViT model. This results in a 3D image with locality in three dimensions. In the original construction of the ViT model, the image is broken into a series of patches of a certain size before being fed into the transformer (after positional encoding). We do something similar but patch over three dimensions instead of two. Specifically, we break our $X$ matrix, which serves as the input, into $(W_{Vbias}, W_{B}, W_{\mu})$ sized windows or patches. The specific size of the patches depends on the neural network result being shown, but most results use (10, 5, 5) sized window segments. We note that $V_\text{Bias}$ has a much larger window than the other measurement parameters. As previously mentioned, we need to generate a larger number of $V_\text{Bias}$ steps to perform the convolution necessary to account for finite temperature effects. This results in additional $V_\text{Bias}$ measurements being generated, whether or not they benefit the machine learning model, simply to achieve an accurate physical simulation. Since we measure many more $V_\text{Bias}$ values compared to the other experimentally tunable parameters, we found it prudent to use a larger $V_{Bias}$ window. The patching is performed using a 3D convolutional layer with a total of 128 filters, which can be considered analogous to the embedding size in NLP (natural language processing). After embedding, the network is fed into a reshaping layer to transform the data into 180 (the total number of patches) by 128 (the size of the embedding) shape. The data then has a simple additive positional encoding applied to it. This encoding is necessary for the multi-head attention mechanism to comprehend the locality of the data, as within the mechanism, the patches are processed in parallel by default, losing the positional information. Following this, the data is fed into a sequence of four vision transformers in series. These layers perform the majority of the computation and account for most of the resource requirements. Within the vision transformer, we use a multi-layer perceptron (MLP) network consisting of one 128-node and one 256-node dense layer, both with a small dropout rate of 0.1 to prevent over-fitting. A diagram of the vision transformer is shown in Fig.\ref{fig:NNDiagram}b. After the four vision transformers, the data is processed by an MLP with two 256-node dense layers and a higher dropout rate of 0.3.

Following this, our network significantly deviates from the standard generalized ViT version. We split the network by incorporating a tree ending or multiple paths at the end of the network. The neural network outputs an array of predictions for our Majorana indicators, with each prediction having its own small MLP perceptron network consisting of two 25-node dense layers. The number of actual paths varies depending on the amount of aggregation within the prediction of the indicators. However, since this is the focus of this paper—namely, the prediction of the full topological phase diagram (or later the phase diagrams of alternative Majorana indicators), we make use of 100 small multi-layer perceptron (dense) networks. This number, 100, corresponds to the number of $B$ measurements multiplied by the number of $\mu$ measurements, where we do not need to include $V_{Bias}$ as the Majorana indicators are only relevant for zero bias. The tree structure is employed because, for a large number of graph points within the output, we require additional, potentially quite complicated, function-based processing for the individual measurement configurations. Using our tree structure is somewhat inefficient, and ideally, in scaled-up versions of our method, it would be preferable to use a convolutional decoder \cite{percebois2023reconstructing}. However, since we are only dealing with 100 points, the tree of MLPs was deemed more than sufficient. {To construct this network we make use of the package Keras \cite{chollet2015keras}, however, we would recommend that future users make use of Pytorch \cite{paszke2019pytorch} instead, which has vision transformers built-in \cite{torchvisionViT}. Those wishing to use our already trained model see \footnote{The authors are happy to help those who contact us to utilize the latest iteration of our model, which is continuously evolving as AI models improve.}.}

In terms of additional details for training, we use Adam's method to optimize the model parameters, adjusting the learning rate from 1E-4 to 1E-6 until no further improvement is observed. We assess the validation of our neural network by splitting our generated data into 90\% for training and setting aside 10\% for testing/validation.  The amount of training data varied slightly for the different parameter regimes discussed; however, approximately 20,000-30,000 training realizations were used in all cases and found to be adequate for analysis. { We trained the network starting from a randomly initiated state without any pre-training. In training we made use of a single A6000, where the time varied significantly depending on the network in no case did it take longer than a few hours with 30 minutes normally being enough for the bulk of convergence.} { The training data generation code is almost identical to that found at \cite{ConductanceKwant}.}

\subsection{Disorder Regimes}
\label{regimes}
To accurately simulate a wide range of potential Majorana devices, and more importantly, to account for current limitations in determining the precise fitting parameters for the nanowire models, we will consider a wide range of possible values. First and foremost, we include a disorder potential generated by impurities in the device, represented by the function $V_{\text{dis}}(x)$. The spatial disorder potential is assumed to vary randomly with position $x$ across the length of the wire with both unknown amplitude and correlation length. This disorder function is the principal problem with Majorana devices and is also generally unknown, i.e. given a superconductor-semiconductor hybrid nanowire device, there is no easy direct way of ascertaining the amount of disorder in it. There are ML methods which as a proof of principle, seem to allow one to determine this disorder function from conductance measurements, they are computationally intensive and difficult to implement in practice for experimental samples \cite{taylor2024machine}. We will assume, as is the case in current experiments, that this disorder is unknown and will consider a variety of different parameter regimes in terms of both its magnitude and correlation length. Beyond this, we will also assume that the spin-orbit coupling constant $\alpha$ is unknown. We note that the real achievement of the current ViT based work is that we do not need to know anything about the minute details of the disorder potential and the spin-orbit coupling in order to figure out if a particular set of conductance measurements in a device implies topological MZMs or not.  Our machine learning algorithm predicts how topological a device is based only the conductance measurements, and we can vary the thresholds used in the indicator diagnostics to make this topology arbitrarily more accurate.

We will consider three disorder regimes, labeled as the {extremes regime}, the full regime, and the moderate regime. The {extremes regime} will consider devices with either high or low disorder. {This is similar to the setup in \cite{cheng2024machine} to allow a direct comparison between our method and previous XGBoost \cite{chen2016xgboost} based methods.} This regime is relatively simple since we expect the device to be topological and trivial respectively in the low and high disorders, and this regime is {'extremes'} because the intermediate regime of moderate disorder is left out. Specifically, the disorder magnitude will be selected from two ranges: [0.15 meV, 1.05 meV] and [3 meV, 4.5 meV], with equal probability of choosing either range and a uniform distribution within each. The disorder correlation length will also be randomly selected from the whole range [20 nm, 70 nm]. Each disorder realization is generated from a Gaussian distribution, with its standard deviation and correlation length chosen from uniform distributions. The correlation length is accounted for by performing a convolution of site-dependent strengths selected from the Gaussian distribution. The full disorder regime, which will be the main focus of our study, is similar to the {extremes regime} but without the division into separate ranges. In this case, the disorder magnitude is randomly selected within the range of [0.15 meV, 4.5 meV]. This regime is expected to be significantly more challenging than the {extremes regime}. The moderate disorder regime, which is likely the most relevant to current experimental devices, has a fixed disorder magnitude of 1.5 meV. It is important to reiterate that this represents the standard deviation of the Gaussian distribution from which the disorder is drawn, rather than a constant disorder function. Note that this disorder characterization is chosen consistent with the recent experiment \cite{aghaee2023inas}.  All earlier experiments had typically an order of magnitude higher disorder, and were all essentially in the very high disorder regime with no topology\cite{das2023search}. The moderate regime is motivated by having the same expected magnitude of disorder in the recent experiment \cite{aghaee2023inas}. Due to being on the boundary between low and high disorder, this moderate disorder regime should be even more challenging for our neural network to classify than the other regimes. In all regimes, the coupling constant $\alpha$ will also be varied randomly, assumed to be within the range [6.8, 9.2], selected from a uniform distribution. Again, the bounds for alpha are chosen based on the physical expectations relevant for the experiment in \cite{aghaee2023inas}.

%May add part with B and mu regime

These regimes are summarized in the table below:

\begin{table}[H]
\begin{tabular}{|c |c |c |c|}
\hline
Regime& Extremes & Full & Moderate\\
 \hline
$\sigma_\text{dis}$(meV) & [0.15, 1.05]+[3, 4.5] & [0.15, 4.5] & [1.5]\\
$L_C$(nm) & [20, 70] & [20, 70] & [20, 70]\\
$\alpha$(nm*meV) & [6.8, 9.2] & [6.8, 9.2] & [6.8, 9.2] \\
$B$(T)& [0.0,0.8] &[0.0,0.8] &[0.0,0.8]\\
$\mu$ (meV)& [0.2,0.4] &[0.2,0.4] &[0.2,0.4]\\
\hline
\end{tabular}
\caption{Summary of parameter ranges for different disorder regimes considered.}
\end{table}

\section{Results}

\subsection{Aggregated Topological Invariant}

In this section, we will present results that most closely align with prior work on this topic, where the goal is to determine from conductance measurements whether a topological phase exists within a device. Specifically, this means identifying whether there is a $B$ and $\mu$ regime within our area of consideration for which $T_V < 0$. This is crucial because, in experimental settings, being able to confidently declare the creation of a topologically non-trivial Majorana nanowire is an incredibly challenging task. Achieving high confidence in such a declaration is of great significance. Additionally, this serves as a useful point of comparison because it closely resembles other methods, such as the topological gap protocol \cite{pikulin2021protocol,pan2021three}, where the accuracy is assessed based on the existence of a single point within a declared topological patch where the scattering topological invariant is negative (independent of how small it may be). In this section, the neural network's goal will be to predict the lowest value of $T_V$ that can be achieved within the range of measurement parameters we simulate (see Section \ref{regimes}).

To begin, we will consider the simplest case: the {extremes regime}, where the goal is to determine whether the device allows a non-trivial topological phase. This regime is the easiest to solve and { serves as a primer for the others. It closely resembles the setup in previous classification work with XGBoost \cite{cheng2024machine}, which distinguished between strongly and weakly disordered wires.} In the {extremes regime}, the neural network is found to have the ability to predict whether a wire has a non-trivial topological phase with near-perfect accuracy. If a device is declared topological by the neural network, there is a 99.79\% probability that a topological phase exists. This accuracy can be further improved by using a slightly more negative cutoff threshold for $T_V$, which we refer to as the passing cutoff $C_\text{{cutoff}}$. This cutoff represents the threshold at which our scheme will declare a device as being topologically non-trivial. Specifically, if the predicted $T_V^\text{Pred}<C_\text{{cutoff}}$, we will declare the device to have passed. The false positive probability ($F_P$) given a topological cutoff is defined as $F_P = P(T_V > 0 | T_V^\text{Pred} < C_\text{{cutoff}})$, or, in other words, the probability that a device declared as passing actually passes is $P(T_V < 0 | \text{Pass}) = 1 - F_P$. 

By using -0.3 as the passing cutoff, we find that within the {extremes regime}, the probability that a wire is topological if declared topological is $P(T_V<0|\text{Pass})=99.93\%$. Using a topological cutoff of -0.6 and within our test data, we do not find a single instance of a device being declared topological that was not. This implies $P(T_V<0|\text{Pass})\geq 99.98\%$, but for our (limited) test data, it shows as $100\%$. We stress that this is achieved using nothing but conductance measurements fed into our neural network—measurements that are routinely done on every nanowire device. One point that may have initially seemed surprising is that we opted for a continuous prediction of $T_V$ instead of a categorical binary yes/no classification, given that the goal is to classify a device as topological or not. While we initially did try this binary classification finding that in terms of predicting whether a device has $T_V<0$, there were very similar fidelities compared to the continuous version, however the continuous version allows much more flexibility in adjusting the cutoff to a desired confidence level { while allowing a physical interpretation}. This is especially important for finite-length wires, where simply having $T_V<0$ may not be sufficient to capture whether a device has the properties of being topologically non-trivial, as an infinite wire would. We believe that the continuous version of $T_V$ used by us is appropriate for the topological diagnosis of real nanowires used experimentally.

\begin{figure}[H]
     \centering
     \begin{subfigure}[b]{0.9\columnwidth}
         \centering
         \includegraphics[width=\textwidth]{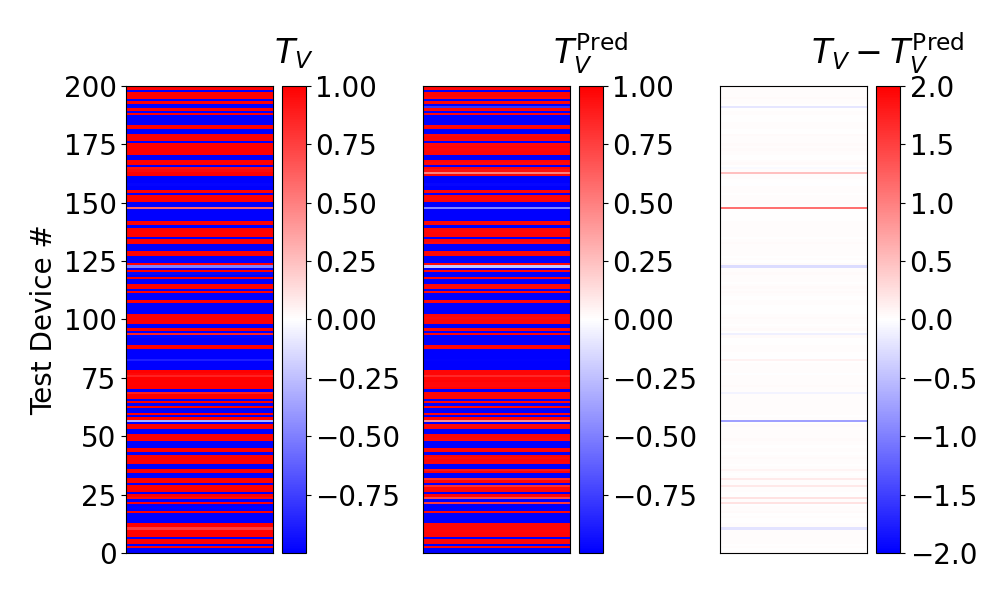}
         \caption{}
     \end{subfigure}
     \begin{subfigure}[b]{0.9\columnwidth}
         \centering
         \includegraphics[width=\textwidth]{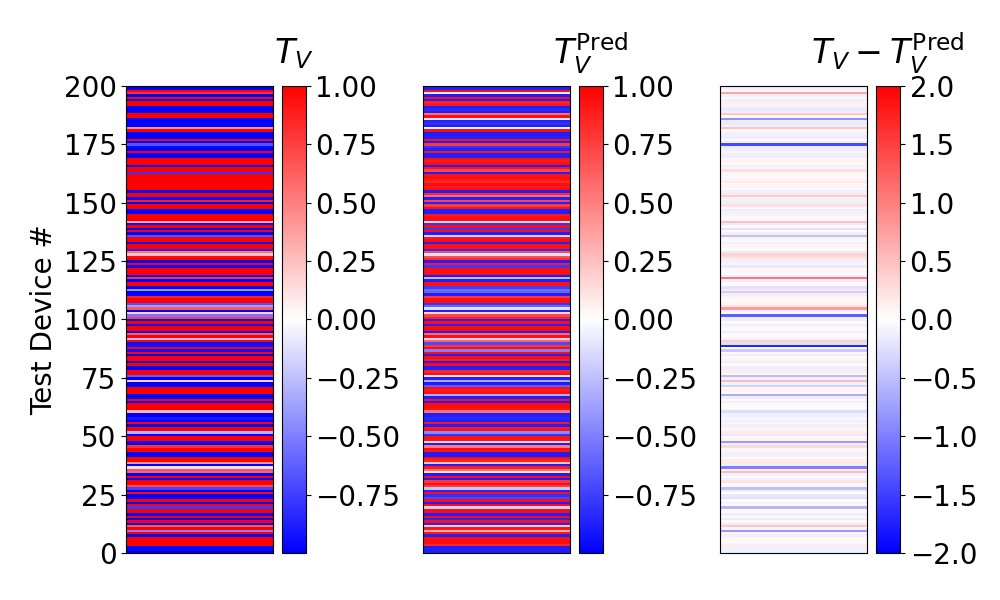}
         \caption{}
     \end{subfigure}
     \begin{subfigure}[b]{0.9\columnwidth}
         \centering
         \includegraphics[width=\textwidth]{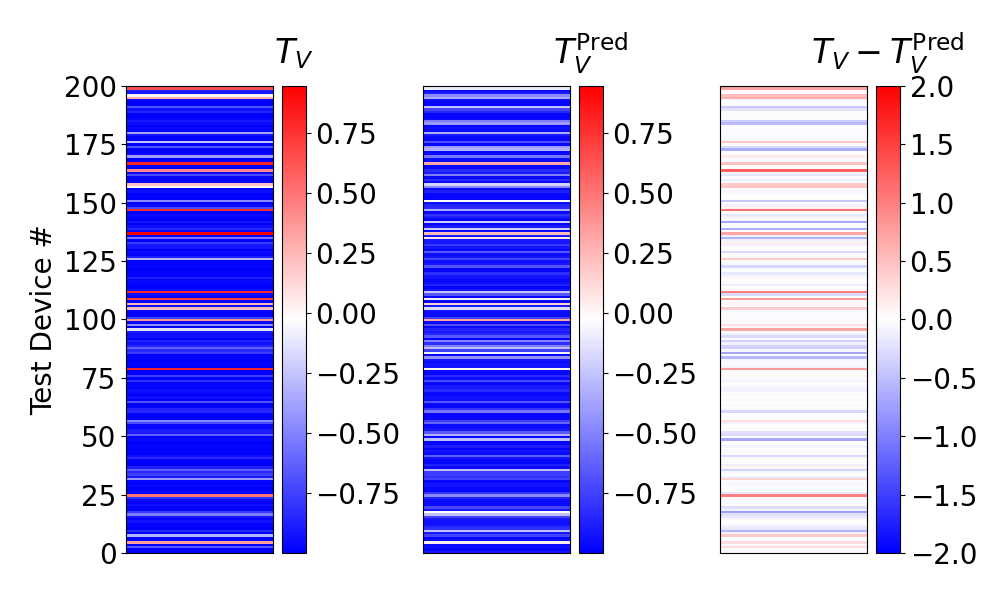}
         \caption{}
     \end{subfigure}
     \caption{Comparisons between the expected minimum topological invariant $T_V$ and the predicted $T_V^\text{Pred}$ by the neural network for different devices. The sub plots are for regimes (a) {extremes regime} (b) full regime (c) moderate regime. The columns are: the expected $T_V$, the predicted $T_V^{Pred}$ from the neural network and the difference between them $T_V-T_V^{Pred}$. Scattering invariants are plotted for 200 representative validation samples with y axis representating the particular device number.}
     \label{fig:TVAggAll}
\end{figure}

One might wonder whether this approach results in discarding a large number of devices—in other words, while the few devices declared topological by the scheme are indeed topological, it is possible that the amount of topological devices that are being incorrectly classified as non-topological increases dramatically. Finding a few 'lucky' devices to be topological simply because a vast majority of the devices is being rejected is not a useful methodology for experimental progress. To explore this, we consider the fraction of devices declared as passing for a given cutoff. When the topological cutoff is set to 0, that fraction is $0.5008$, which is unsurprising given the {extremes regime} involves randomly selecting between high or low disorder, and hence roughly half the devices are topological/trivial. When the cutoff is tightened to $-0.5$, the fraction of devices declared topological only drops slightly to $0.4972$, indicating that almost no topological devices are being lost in the {extremes regime}. The exact value of the topological invariant for the most negative point in a device could be determined with a root mean square error of $R_{MS}(T_V)=0.1473$. For a figure comparing the predicted and expected topological invariants in the aggregated {extremes regime}, see Fig. \ref{fig:TVAggAll}a.
%-------------
These results can be seen summarized in the table below clearly showing the high efficacy of our machine learning algorithm:
\begin{table}[H]
\begin{tabular}{|c |c |c |c|}
\hline
Cutoff	&$1-F_P$	&$1-F_N$	&P(Passing)\\
 \hline
0.0000&0.9952&0.9871&0.5008\\
-0.3000&0.9976&0.9888&0.4988\\
-0.5000&0.9992&0.9912&0.4972\\
-0.7000&1.0000&0.9920&0.4964\\
-0.8000&1.0000&0.9936&0.4948\\
-0.9000&1.0000&0.9960&0.4924\\
\hline
\end{tabular}
\caption{Summary of $T_V$ confidence results for the fully aggregated {extremes regime}. Results are shown for different cutoff values for which a device will be declared passing. $F_P$ is the false positive probability, which implies $1-F_P$ is the probability that a device declared passing, has a $T_V<0$. $F_N=P(T_V<C_{cutoff}|T_V^{Pred}>0)$ is the false negative probability. P(Passing) describes the fraction of devices being declared passing the scheme, namely devices with $T_V^{Pred}<C_{cutoff}$. The root mean squared error between the expected and predicted results was $R_{MS}(T_V)=0.1473$.}
\end{table}

Moving on, the {extremes regime}, while most similar to previous results, is unrealistic since it only considers very high or very low disorder. In reality, devices have a full continuum of disorder, with some exhibiting the more challenging moderate disorder. When examining the full regime, we observe a realistic (and expected) drop in the fidelity of our neural network's ability to determine whether a device is topological. Specifically, we find that for the full regime, the probability $P(T_V<0|\text{Pass})=0.9461$. Although this represents a significant drop, it can be mitigated by increasing the topological cutoff. We find that for a topological cutoff of $-0.3$, $P(T_V<0|\text{Pass})$ increases significantly to $0.9903$, and when the $C_{cutoff}$ is further lowered to $-0.8$, no devices in our test data are incorrectly declared topological, implying a probability of a topological pass being accurate at a fidelity level of $>0.9998$. Of course, there is a trade-off with lowering the cutoff, as it leads to more topological devices being missed and an associated higher rate of rejections. However, for a cutoff of $-0.3$, the reduction is very marginal, being $0.4576$ at $C_{cutoff}=0$ dropping to $0.4058$ at $C_{cutoff}=-0.3$. For $C_{cutoff}=-0.8$, the drop is larger, with the fraction passing becoming $0.3077$, but still relatively acceptable. This demonstrates that, within the full regime, one can achieve an arbitrarily high level of confidence in a device being topologically non-trivial by adjusting the topological cutoff while at the same time keeping the device rejection rate reasonably small. The exact value of the topological invariant for the most negative point in a device can be determined with an error of $R_{MS}(TV) = 0.3134$. These results are summarized in the table below and are shown in Fig. \ref{fig:TVAggAll}b:

\begin{table}[H]
\begin{tabular}{|c |c |c |c|}
\hline
Cutoff	&$1-F_P$	&$1-F_N$	&P(Passing)\\
 \hline
0.0000&0.9461&0.9400&0.4576\\
-0.3000&0.9903&0.9573&0.4058\\
-0.5000&0.9961&0.9691&0.3787\\
-0.7000&1.0000&0.9764&0.3412\\
-0.8000&1.0000&0.9845&0.3077\\
-0.9000&1.0000&0.9918&0.0010\\
\hline
\end{tabular}
\caption{Summary of $T_V$ confidence results for the fully aggregated full disorder regime.  Results are shown for different cutoff values for which a device will be declared passing. $F_P$ is the false positive probability, which implies $1-F_P$ is the probability that a device declared passing, has a $T_V<0$. $F_N=P(T_V<C_{cutoff}|T_V^{Pred}>0)$ is the false negative probability. P(Passing) describes the fraction of devices being declared passing the scheme, namely devices with $T_V^{Pred}<C_{cutoff}$. The root mean squared error between the expected and predicted results was $R_{MS}(T_V)=0.3134$.}
\end{table}

Next we discuss the moderate disorder regime, where it is difficult to determine whether a device is in a topological phase. At a topological $C_{cutoff}=0$, the probability $P(T_V<0|\text{Pass})=0.9634$, which seems to represent an increase from the full regime. However, this is a byproduct of the fraction passing being $0.95$ as in almost all devices in this disorder regime have 1 point on their phase diagram with $T_V<0$. This fidelity can be increased $P(T_V<0|\text{Pass})>0.9998$ by tuning the $C_{cutoff}$ to -0.9 though with a 38\% reduction in devices passing. Looking at $R_{MS}(T_V)=0.2659$ shows a slightly smaller error in $T_V$ prediction. Going further, since this is the most experimentally relevant regime, a $T_V<0$ may be insufficient to determine whether short disordered devices have useful topological properties. We therefore introduce an expectation cutoff $C^{expect}_{cutoff}$.  It is straight forward to consider then $P(T_V<C^\text{expect}_{cutoff}|T_V^\text{Pred}<C_{cutoff}])$. When we are trying to achieve a $T_V<-0.5$ the probability that a device declared passing has the required $T_V$ is $P(T_V<-0.5|T_V^\text{Pred}<-0.5)=0.9612$. Once again by tuning the $C_{cutoff}$ to -0.8, the fidelity increases to $P(T_V<-0.5|Pass)=0.9929$, this is despite only a 16\% reduction in the percent of devices passing. For $T_V<-0.9$ and $C_{cutoff}=-0.9$ the $P(T_V<-0.9|Pass)=0.9558$ lowering the $C_{cutoff}$ to -0.95, this can be increased to 0.9875 though with a 30\% reduction in the number of devices passing. These results suggest that not only can our scheme determine with high confidence whether a device has a topological phase, it can also determine whether a device has a strongly passing $T_V$. Although the fraction of passing devices deceases with increasing $T_V$ cut off threshold, a reasonable balance between determining topology and an acceptable number of passing devices seems possible. These results are summarized in the table below and can be seen in Fig. \ref{fig:TVAggAll}c:

\begin{table}[H]
\begin{tabular}{|c |c |c |c|}
\hline
$C_{cutoff}$	&$P(T_V<C^\text{expect}_{cutoff}|T_V^\text{Pred}<C_{cutoff}])$	&P(Passing)\\
 \hline
 &$C^\text{expect}_{cutoff}=0$ &   \\
 \hline
0.0000&0.9634&0.9551\\
-0.3000&0.9821&0.9051\\
-0.5000&0.9902&0.8632\\
-0.7000&0.9977&0.7824\\
-0.8000&0.9990&0.7224\\
-0.9000&1.0000&0.5908\\
\hline
&$C^\text{expect}_{cutoff}=-0.5$ &   \\
 \hline
-0.5000&0.9612&0.8632\\
-0.7000&0.9826&0.7824\\
-0.8000&0.9929&0.7224\\
-0.9000&0.9994&0.5908\\
-0.9500&1.0000&0.4132\\
\hline
&$C^\text{expect}_{cutoff}=-0.9$ &   \\
 \hline
-0.9000&0.9558&0.5908\\
-0.9500&0.9875&0.4132\\
\hline
\end{tabular}
\caption{Summary of $T_V$ confidence results for the fully aggregated moderate disorder regime. Results are shown for different cutoff values for which a device will be declared passing. $P(T_V<C^\text{expect}_{cutoff}|T_V^\text{Pred}<C_{cutoff})$ is the probability that a device declared passing has a $T_V<C_{cutoff}^\text{expect}$. P(Passing) describes the fraction of devices being declared passing the scheme. The root mean squared error between the expected and predicted results was $R_{MS}(T_V)=0.2659$.}
\end{table}

%Add text about alternative cutoffs. 

\subsection{Continuous Topological Invariant}

From the previous section, we have shown the ability to determine with remarkable accuracy whether a device has a non-topological phase. The next question that naturally arises from this is where precisely in the parameter space the topology may exist. Specifically, can we generate the full phase diagram, which would show exactly the parameter regime needed to make the 1D nanowire topologically non-trivial? Furthermore, we shift from merely assigning a device as topological or not, to also addressing the degree to which it is topologically invariant. In an infinite system, this question is nonsensical since $T_V$ is either positive or negative; however, in a short, finite wire with disorder, the actual $T_V$ magnitude matters.

We seek to determine the quantitative topological phase diagrams of $T_V$ plotted for different values of $\mu$ and $B$. This is important because, to use the 1D nanowire computationally, it is necessary to set these parameters to values where the wire is in a topologically non-trivial state. To provide a qualitative sense of our neural network's ability to predict $T_V$, and to offer insight into how the neural network may fail, we present several figures comparing the expected phase diagram with the neural network's predictions. We, of course, know the expected phase diagram since we can directly calculate $T_V$ for each set of system parameters used in our simulations

To start, we once again consider the {extremes regime}. In this regime, we found that the average root mean square error for $T_V$ was $R_{ms}(T_V)=0.2280$, indicating a high fidelity. This value represents the average error across all points within the phase diagram. As shown in Fig. \ref{fig:TVBrokeC}, the neural network is remarkably effective at predicting the $T_V$ phase diagram, with errors mainly appearing near expected small red patches within the blue topologically non-trivial bulk or at the phase boundaries. The network also captures the existence of these boundaries by reporting intermediate $T_V$ values between the two phases. Unlike the more chaotic simulated true $T_V$, the neural network's prediction exhibits a smoother transition. The strength of this prediction becomes even more apparent when considering the false positive probabilities. On average, for all points, $P(T_V<0|\text{Pass}) = 0.9164$ when the cutoff is set to 0. Lowering the cutoff to -0.5 increases $P(T_V<0|\text{Pass})$ to $0.9720$, while the fraction of points declared topological drops from 0.2377 to 0.1848. Unlike in the aggregated case, while it is still possible to achieve seemingly arbitrary confidence in a point being declared topologically non-trivial, there is a significant drop-off in the number of devices that satisfy the criteria. It is important to note that these metrics are averaged over all points, so we should expect lower fractions of points to be declared topological, with certain areas of the phase diagram being easier to predict than others. This makes graphical analysis essential for understanding the neural network's errors.

A summary of these results is provided in the tables below:

\begin{table}[H]
\begin{tabular}{|c |c |c |c|}
\hline
Cutoff	&$1-F_P$	&$1-F_N$	&P(Passing)\\
 \hline
0.0000&0.9164&0.9817&0.2377\\
-0.3000&0.9447&0.9860&0.2123\\
-0.5000&0.9720&0.9885&0.1848\\
-0.7000&0.9853&0.9913&0.1307\\
-0.8000&0.9925&0.9930&0.0929\\
-0.9000&0.9968&0.9951&0.0338\\
-0.9500&0.9996&0.9966&0.0091\\
\hline
\end{tabular}
\caption{Summary of $T_V$ confidence results for the {extremes regime}. Results are shown for different cutoff values for which a device will be declared passing. $F_P$ is the false positive probability, which implies $1-F_P$ is the probability that a device declared passing, has a $T_V<0$. $F_N=P(T_V<C_{cutoff}|T_V^{Pred}>0)$ is the false negative probability. P(Passing) describes the fraction of phase points having $T_V^{Pred}<C_{cutoff}$. The root mean squared error between the expected and predicted results is $R_{MS}(T_V)=0.2492$.}
\end{table}

Moving onto the full regime, we expect the phase diagrams to be more complex and thus more difficult to predict. This expectation is exactly what we observe. In the full regime, the phase diagrams become more intricate in shape, but despite this, the neural network manages to predict their shape with high accuracy. The error in the full regime only slightly increases to $R_{MS}(T_V)=0.2687$, and is shown in Fig. \ref{fig:TVFullC}, there is a strong quantitative agreement between the predicted and expected $T_V$. In this full regime, which is the most important regime, the neural network, based solely on a sequence of conductance measurements, can accurately predict the entire topological phase diagram. This holds true even when large, complex features are present in the phase diagram. When the neural network fails to capture certain characteristics, it does so in a predictable manner, often producing a phase diagram that appears as if it had undergone some form of Gaussian filtering. The probability $P(T_V<0|\text{Pass})$ for a point on the phase diagram drops to 0.8687, but by lowering the cutoff, this loss can be recovered, allowing for confidence to be restored to near-perfect levels. 

A summary of these results is provided in Table \ref{table:fulltv} below:

\begin{table}[H]
\begin{tabular}{|c |c |c |c|}
\hline
Cutoff	&$1-F_P$	&$1-F_N$	&P(Passing)\\
 \hline
0.0000&0.8687&0.9789&0.1304\\
-0.3000&0.9270&0.9847&0.1011\\
-0.5000&0.9502&0.9879&0.0788\\
-0.7000&0.9702&0.9915&0.0491\\
-0.8000&0.9784&0.9936&0.0313\\
-0.9000&0.9818&0.9957&0.0092\\
-0.9500&0.9840&0.9972&0.0034\\
\hline
\end{tabular}

\caption{Summary of $T_V$ confidence results for the full regime. Results are shown for different cutoff values for which a device will be declared passing. $F_P$ is the false positive probability, which implies $1-F_P$ is the probability that a device declared passing, has a $T_V<0$. $F_N=P(T_V<C_{cutoff}|T_V^{Pred}>0)$ is the false negative probability. P(Passing) describes the fraction of phase points having $T_V^{Pred}<C_{cutoff}$. The root mean squared error between the expected and predicted results is $R_{MS}(T_V)=0.2687.$ { The false negative rate seems to decrease due to a side effect of averaging over a device where most points with $T^{Pred}_V>0$ are obviously non-topological, while whether the most negative $T_V$ point on a device is non-topological is often not. The opposite is true for the false positive rate, as the shape of topology is often difficult, but whether a device possess a topological region is normally much easier.} }
\label{table:fulltv}
\end{table}

\begin{figure}[H]
     \centering
     \begin{subfigure}[b]{0.9\columnwidth}
         \centering
         \includegraphics[width=\textwidth]{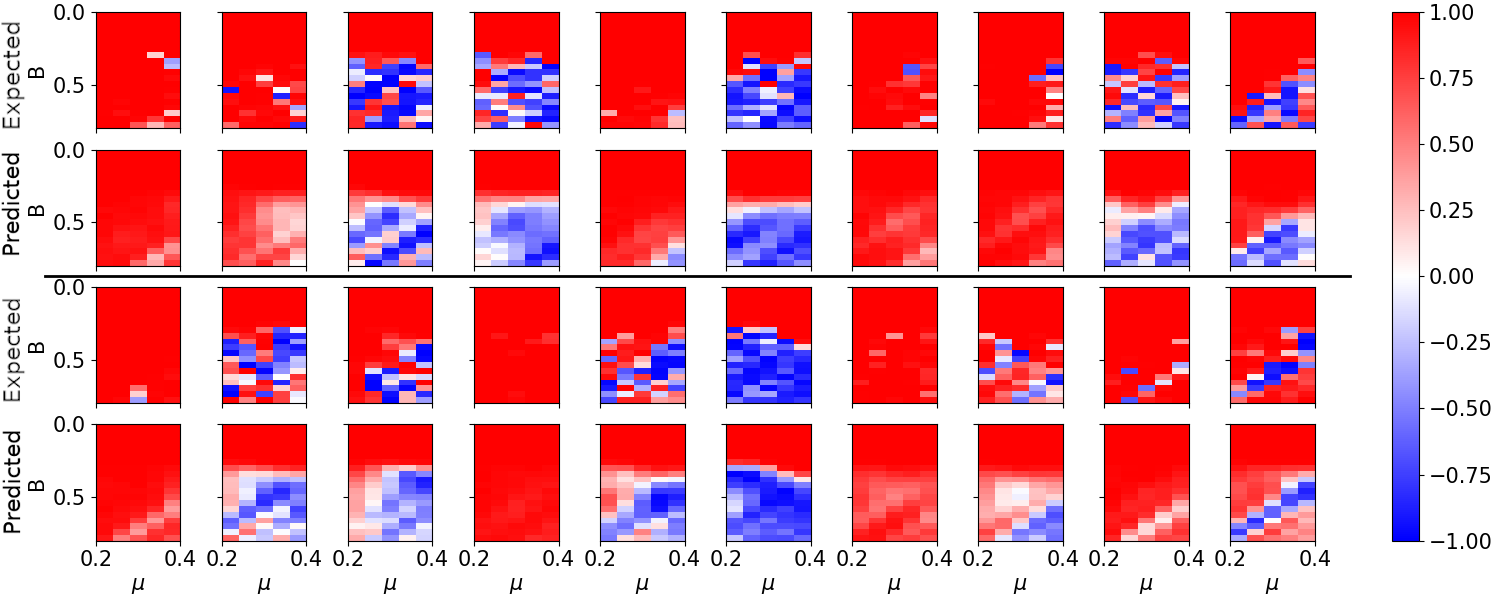}
         \caption{}
     \end{subfigure}
     \caption{Comparison between expected $T_V$ and predicted $T_V^{Pred}$ phase diagrams for the moderate regime. Plots are arranged in vertical pairs where the top plot is the expected $T_V$ and the bottom is the predicted $T_V^{Pred}$. In all plots the phase diagrams are plotted for magnetic field $B$ vs. chemical potential $\mu$ with a continuous $T_V$. Each pair is a different representative sample from the validation set selected at random.}
     \label{fig:TVModC}
\end{figure}

The moderate regime brings far more complexity to the problem. Its phase diagram is much more complicated than that of the full regime, as almost the entirety of the data has complicated and intricate features that are difficult to predict or explain. This is true both for the neural network and for manual analysis. For example, the reported phase diagrams in Ref. \cite{aghaee2023inas} are extremely complex with patchy topological phases often surrounded by large trivial phases. Despite these difficulties, Fig. \ref{fig:TVModC} shows that our model is still able to accurately capture the complex features of the phase diagrams, even though the error $R_{MS}(T_V)=0.4049$ increases quite significantly. One possible explanation for the increased error is that the phase diagram has more boundaries between phases due to the increased complexity. The neural network's ability to predict bulk regions of $T_V$ does not seem to be significantly lower, with virtually no qualitative difference observed. Once again, this supports the idea that the errors in the neural network's predictions resemble a Gaussian filtering effect. Given this, and considering that in practice one needs a point within the phase diagram where one can confidently determine that the device is topologically non-trivial, it may be optimal to use a Gaussian-filtered version (with a low-strength filter) of the neural network output when selecting the point of optimal $T_V$. This approach is beneficial because the neural network can be used to predict the blended version, and when blending, the lowest $T_V$ will not become lower. %As long as one stays strongly within the bulk region in the blended phase diagram they can be confident they have a topologically non-trivial device. 

%-Talk about how it seems to predict the result but not with the full crispness of the the expected value. The neural network seems to provide a phase diagram as if there was some Gaussian blending.

A summary of results for the moderate regime is provided in Table \ref{table:modtv}:

\begin{table}[H]
\begin{tabular}{|c |c |c |c|}
\hline
Cutoff	&$1-F_P$	&$1-F_N$	&P(Passing)\\
 \hline
0.0000&0.7623&0.9461&0.2015\\
-0.3000&0.8539&0.9607&0.1327\\
-0.5000&0.9089&0.9697&0.0876\\
-0.7000&0.9449&0.9791&0.0409\\
-0.8000&0.9617&0.9839&0.0203\\
-0.9000&0.9722&0.9898&0.0064\\
-0.9500&0.9753&0.9939&0.0030\\
\hline
\end{tabular}
\caption{Summary of $T_V$ confidence results for the moderate regime. Results are shown for different cutoff values for which a device will be declared passing. $F_P$ is the false positive probability, which implies $1-F_P$ is the probability that a device declared passing, has a $T_V<0$. $F_N=P(T_V<C_{cutoff}|T_V^{Pred}>0)$ is the false negative probability. P(Passing) describes the fraction of phase points having $T_V^{Pred}<C_{cutoff}$. The root mean squared error between the expected and predicted results is $R_{MS}(T_V)=0.4049$.}
\label{table:modtv}
\end{table}

\clearpage
\onecolumngrid
\begin{minipage}{\linewidth} 
\begin{figure}[H]
     \centering
     \begin{subfigure}[b]{0.90\columnwidth}
         \centering
         \includegraphics[width=\textwidth]{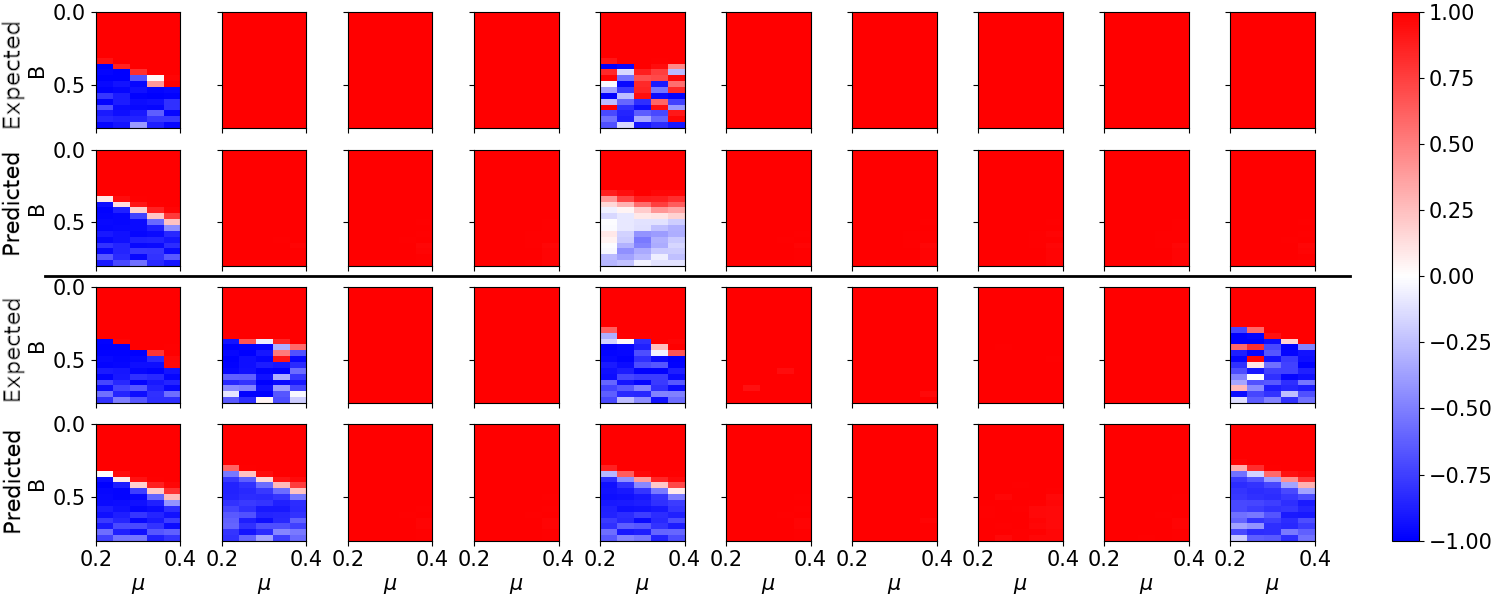}
         \caption{}
     \end{subfigure}
     \caption{Comparison between expected $T_V$ and predicted $T_V^{Pred}$ phase diagrams for the {extremes regime}. Plots are arranged in vertical pairs where the top plot is the expected $T_V$ and the bottom is the predicted $T_V^{Pred}$. In all plots the phase diagrams are plotted for magnetic field $B$ vs. chemical potential $\mu$ with a continuous $T_V$. Each pair is a different representative sample from the validation set selected at random.  }
     \label{fig:TVBrokeC}
\end{figure}
\begin{figure}[H]
     \centering
     \begin{subfigure}[b]{0.95\columnwidth}
         \centering
         \includegraphics[width=\textwidth]{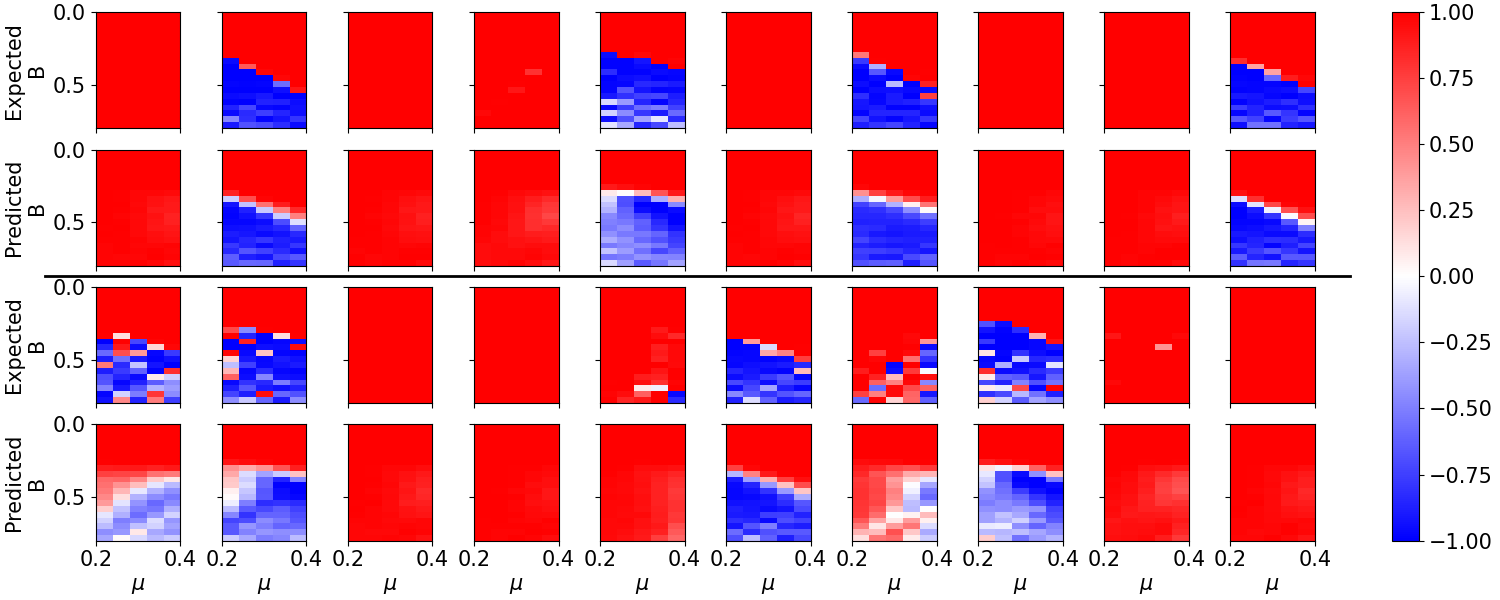}
         \caption{}
     \end{subfigure}
     \caption{Comparison between expected $T_V$ and predicted $T_V^{Pred}$ phase diagrams for the full regime. Plots are arranged in vertical pairs where the top plot is the expected $T_V$ and the bottom is the predicted $T_V^{Pred}$. In all plots the phase diagrams are plotted for magnetic field $B$ vs. chemical potential $\mu$ with a continuous $T_V$. Each pair is a different representative sample from the validation set selected at random. }
     \label{fig:TVFullC}
\end{figure}

\end{minipage}
\clearpage

\twocolumngrid

\subsection{Continuous Alternative Indicators}

So far, our analysis has been based entirely on the scattering invariant or its continuous version $T_V$. Moving onto the alternative indicators, we only consider continuous generation of their respective phase diagrams. Predicting the phase diagrams for these alternative indicators turns out to be much more challenging than for the topological invariant. The difficulty seems to stem from their inherently more unstable nature, particularly with $I_1$, which relies on the maximum of the LDOS rather than an average in its calculation (the maximum effect does not seem to have the same effect in the aggregated $T_V$ because many points on the plot have approximately the same value and the maximum is thus ill-defined from the perspective of the neural network). We will first consider the indicator $I_2$ within the same relevant regimes discussed earlier, excluding the {extremes regime}, as it was only considered for comparison with previous methods. There has been no prior attempt to learn these alternative indicators, either through machine learning or protocols similar to the topological gap protocol. In this ML protocol, we will, similar to the previous section, apply a cutoff for the prediction of the indicators, below which a device will be declared as passing the indicator test. The topological indicator $I_2$ was predicted within the full regime with an $R_{MS}(I_2) = 0.4640$, indicating a significantly higher difficulty compared to $T_V$. Despite this, the phase diagram shown in Fig. \ref{fig:I_2FullC} still provides strong qualitative agreement, with similar properties to the $T_V$ case. {We should note that these alternative indicators behave quite differently from $T_V$ and cannot be used as a substitute for it; rather, they should be used in conjunction with $T_V$. This is because these indicators only determine whether the LDOS is localized on the edge without domain walls, not whether that localization is topological. Trivial Andreev bound states often satisfy this condition, and thus, it will commonly appear in diagrams as if these indicators pass in almost opposite regions to $T_V$. However, this is topologically meaningless, as these indicators imply nothing without $T_V$ passing as well.} We find that at $C_{cutoff}=0$, the probability that a device phase point declared passing has $I_2 < 0$ is $P(I_2 < 0|\text{Pass}) = 0.8780$. By lowering the cutoff, one can significantly improve the accuracy of a device phase point being correctly declared as passing; however, unlike in the $T_V$ case, the fraction of phase points considered passing significantly decreases as the cutoff is lowered. At a mixed point of $C_{cutoff} = -0.5$, $P(I_2 < 0|\text{Pass})$ increases to $0.9476$, but the fraction of setups passing drops from $0.6391$ to $0.4331$. For a full summary of these results, see the Table \ref{table:i2full} below:

\begin{table}[H]
\label{table:I2}
\begin{tabular}{|c |c |c |c|}
\hline
Cutoff	&$1-F_P$	&$1-F_N$	&P(Passing)\\
 \hline
0.0000&0.8780&0.7853&0.6391\\
-0.3000&0.9292&0.8682&0.5280\\
-0.5000&0.9476&0.9189&0.4331\\
-0.7000&0.9639&0.9721&0.3505\\
-0.8000&0.9685&0.9878&0.3003\\
-0.9000&0.9747&0.9965&0.2391\\
-0.9500&0.9895&0.9990&0.1671\\
\hline
\end{tabular}
\caption{Summary of $I_2$ confidence results for the full regime. Results are shown for different cutoff values for which a device will be declared passing. $F_P$ is the false positive probability, which implies $1-F_P$ is the probability that a device declared passing, has a $I_2<0$. $F_N=P(I_2<C_{cutoff}|I_2^{Pred}>0)$ is the false negative probability. P(Passing) describes the fraction of phase points having $I_2<C_{cutoff}$. The root mean squared error between the expected and predicted results is $R_{MS}(I_2)=0.4640$}
\label{table:i2full}
\end{table}

In the moderate regime, interestingly, predicting the $I_2$ indicator does not seem to significantly increase in difficulty, with an $R_{MS}(I_2) = 0.4799$. Similar qualitative agreement between the real and predicted phase diagrams for this regime can be seen in Fig. \ref{fig:I_2ModC}. A summary of these results is provided in Table \ref{table:i2mod}:

\begin{table}[H]
\begin{tabular}{|c |c |c |c|}
\hline
Cutoff	&$1-F_P$	&$1-F_N$	&P(Passing)\\
 \hline
0.0000&0.7267&0.8471&0.4307\\
-0.3000&0.8555&0.9066&0.3288\\
-0.5000&0.9417&0.9415&0.3075\\
-0.7000&0.9839&0.9700&0.2852\\
-0.8000&0.9935&0.9905&0.2679\\
-0.9000&0.9987&0.9977&0.2331\\
-0.9500&0.9983&0.9996&0.1711\\
\hline
\end{tabular}
\caption{$I_2$ Confidence Results for the moderate regime. Results are shown for different cutoff values for which a device will be declared passing. $F_P$ is the false positive probability, which implies $1-F_P$ is the probability that a device declared passing, has a $I_2<0$. $F_N=P(I_2<C_{cutoff}|I_2^{Pred}>0)$ is the false negative probability. P(Passing) describes the fraction of phase points having $I_2<C_{cutoff}$. The root mean squared error between the expected and predicted results is $R_{MS}(I_2)=0.4799$.}
\label{table:i2mod}
\end{table}

Moving onto the $I_1$ indicator, predicting this indicator using the neural network proved to be much more challenging than $I_2$. In the full regime, $R_{MS}(I_1) = 0.4931$, which is a reduction compared to that of $I_2$. $P(I_1 < 0|\text{Pass}) = 0.9074$ is higher even for a $C_{cutoff} = 0$, though this is likely because the probability of passing is already high. Quantitative agreement between the predicted and expected phase diagrams also persists, as shown in Fig. \ref{fig:I_1FullC}. A summary of these results can be seen below in Table \ref{table:i1full} below:

\begin{table}[H]
\begin{tabular}{|c |c |c |c|}
\hline
Cutoff	&$1-F_P$	&$1-F_N$	&P(Passing)\\
 \hline
0.0000&0.9074&0.7419&0.8270\\
-0.3000&0.9394&0.7881&0.7476\\
-0.5000&0.9563&0.8231&0.6925\\
-0.7000&0.9698&0.8671&0.6217\\
-0.8000&0.9762&0.8925&0.5714\\
-0.9000&0.9865&0.9221&0.4760\\
-0.9500&0.9931&0.9406&0.3603\\
\hline
\end{tabular}
\caption{Summary of $I_1$ confidence results for the full regime. Results are shown for different cutoff values for which a device will be declared passing. $F_P$ is the false positive probability, which implies $1-F_P$ is the probability that a device declared passing, has a $I_1<0$. $F_N=P(I_1<C_{cutoff}|I_1^{Pred}>0)$ is the false negative probability. P(Passing) describes the fraction of phase points having $I_1<C_{cutoff}$. The root mean squared error between the expected and predicted results is $R_{MS}(I_1)=0.4931$.}
\label{table:i1full}
\end{table}

 In the moderate regime, the error for $I_1$ increases significantly to $R_{MS}(I_1)=0.6162$. This is because $I_1$ is more unstable and, therefore, harder to predict. This increased instability leads to greater error, as the neural network struggles more at the boundaries of phases, making the Gaussian filter-like prediction much more challenging. Despite this, quantitative agreement, as shown in Fig. \ref{fig:I_1ModC}, remains. However, the $P(I_1 < 0|\text{Pass}) = 0.7448$ at $C_{cutoff} = 0$ is much lower, and can not be enhanced without a significant loss in the number of passing device phase points. To achieve $P(I_1 < 0|\text{Pass}) = 0.9982$, the cutoff must be lowered to -0.8, which results in nearly a 50\% reduction in the number of phase points passing the protocol. A summary of these results is provided in the Table \ref{table:i1mod} below:

\begin{table}[H]
\begin{tabular}{|c |c |c |c|}
\hline
Cutoff	&$1-F_P$	&$1-F_N$	&P(Passing)\\
 \hline
0.0000&0.7448&0.7974&0.6279\\
-0.3000&0.8187&0.8411&0.4836\\
-0.5000&0.8818&0.8752&0.3962\\
-0.7000&0.9682&0.9142&0.3342\\
-0.8000&0.9921&0.9365&0.3184\\
-0.9000&0.9982&0.9677&0.3047\\
-0.9500&0.9985&0.9822&0.3008\\
\hline
\end{tabular}
\caption{Summary of $I_1$ confidence Results for the moderate regime. Results are shown for different cutoff values for which a device will be declared passing. $F_P$ is the false positive probability, which implies $1-F_P$ is the probability that a device declared passing, has a $I_1<0$. $F_N=P(I_1<C_{cutoff}|I_1^{Pred}>0)$ is the false negative probability. P(Passing) describes the fraction of phase points having $I_1<C_{cutoff}$. The root mean squared error between the expected and predicted results is $R_{MS}(I_1)=0.6162$.}
\label{table:i1mod}
\end{table}

%Comparison between expected $T_V$ and predicted $T_V^{Pred}$ phase diagrams for the {extremes regime}. Plots are arranged in vertical pairs where the top plot is the expected $T_V$ and the bottom is the predicted $T_V^{Pred}$. In all plots the phase diagrams are plotted for magnetic field $B$ vs. chemical potential $\mu$ with a continuous $T_V$. Each pair is a different representative sample from the validation set selected at random. 

\clearpage
\onecolumngrid
\begin{minipage}{\linewidth} 

\begin{figure}[H]
     \centering
     \begin{subfigure}[b]{0.95\columnwidth}
         \centering
         \includegraphics[width=\textwidth]{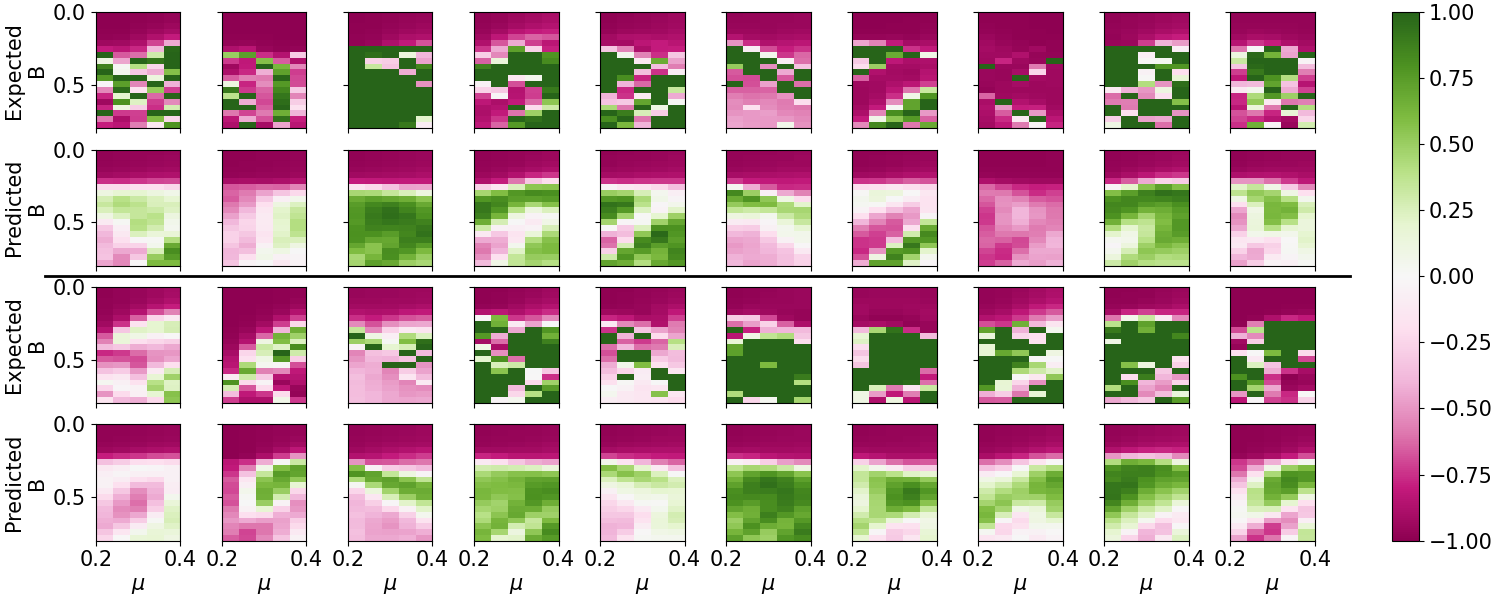}
         \caption{}
     \end{subfigure}
     \caption{Comparison between expected $I_2$ and predicted $I_2^{Pred}$ phase diagrams for the full regime. Plots are arranged in vertical pairs where the top plot is the expected $I_2$ and the bottom is the predicted $I_2^{Pred}$. In all plots the phase diagrams are plotted for magnetic field $B$ vs. chemical potential $\mu$ with a continuous $I_2$. Each pair is a different representative sample from the validation set selected at random. }
     \label{fig:I_2FullC}
\end{figure}
\begin{figure}[H]
     \centering
     \begin{subfigure}[b]{0.95\columnwidth}
         \centering
         \includegraphics[width=\textwidth]{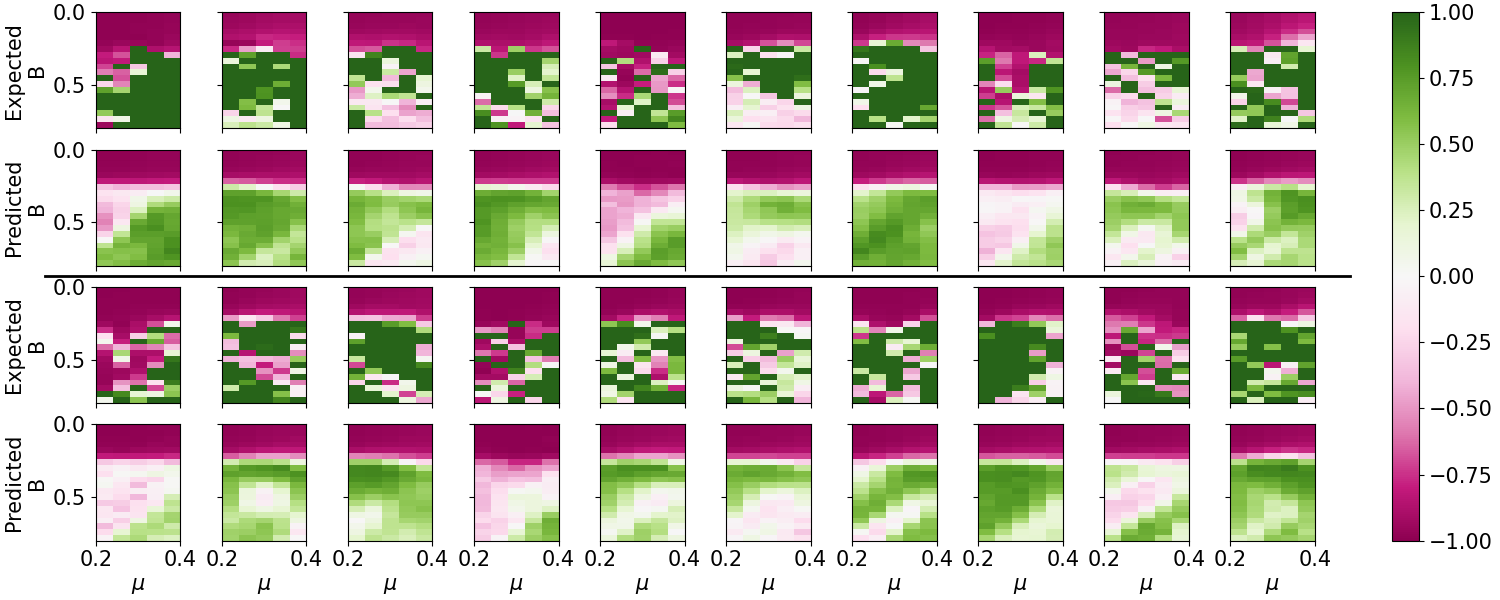}
         \caption{}
     \end{subfigure}
     \caption{Comparison between expected $I_2$ and predicted $I_2^{Pred}$ phase diagrams for the moderate regime. Plots are arranged in vertical pairs where the top plot is the expected $I_2$ and the bottom is the predicted $I_2^{Pred}$. In all plots the phase diagrams are plotted for magnetic field $B$ vs. chemical potential $\mu$ with a continuous $I_2$. Each pair is a different representative sample from the validation set selected at random.}
     \label{fig:I_2ModC}
\end{figure}

\end{minipage}
\clearpage
\twocolumngrid

\clearpage
\onecolumngrid
\begin{minipage}{\linewidth}

\begin{figure}[H]
     \centering
     \begin{subfigure}[b]{0.95\columnwidth}
         \centering
         \includegraphics[width=\textwidth]{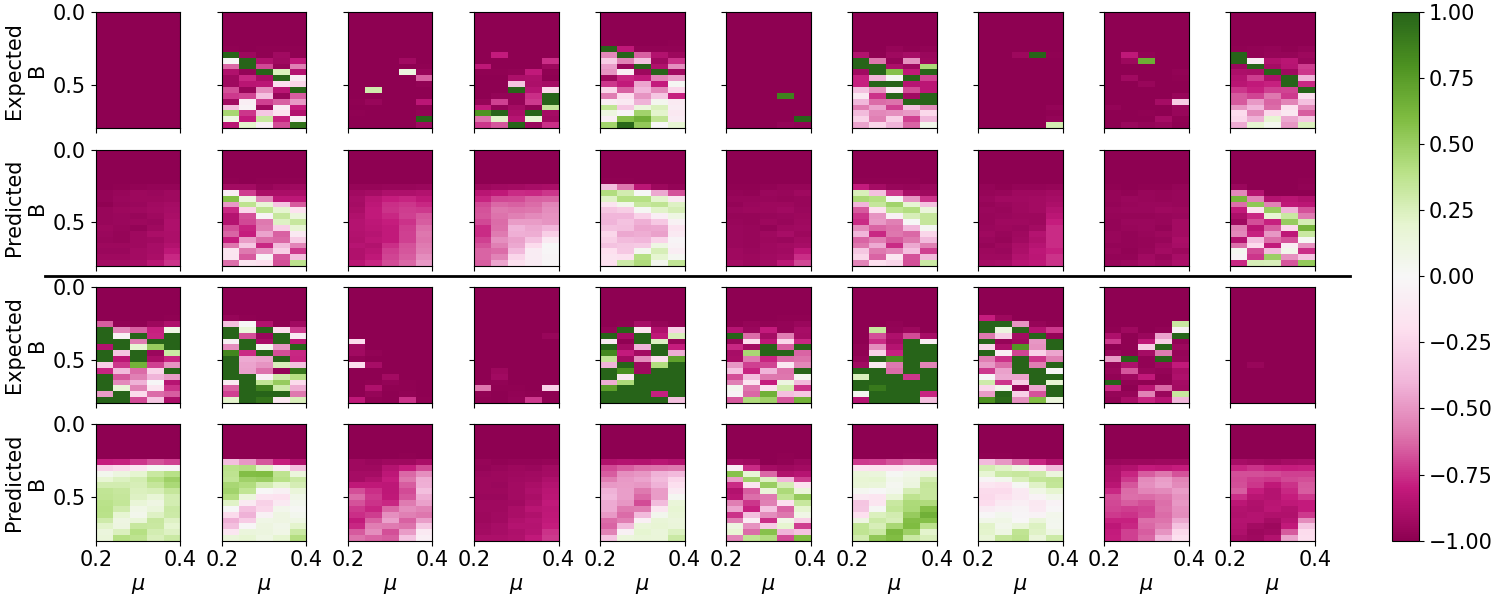}
         \caption{}
     \end{subfigure}
     \caption{Comparison between expected $I_1$ and predicted $I_1^{Pred}$ phase diagrams for the full regime. Plots are arranged in vertical pairs where the top plot is the expected $I_1$ and the bottom is the predicted $I_1^{Pred}$. In all plots the phase diagrams are plotted for magnetic field $B$ vs. chemical potential $\mu$ with a continuous $I_1$. Each pair is a different representative sample from the validation set selected at random. }
     \label{fig:I_1FullC}
\end{figure}
\begin{figure}[H]
     \centering
     \begin{subfigure}[b]{0.95\columnwidth}
         \centering
         \includegraphics[width=\textwidth]{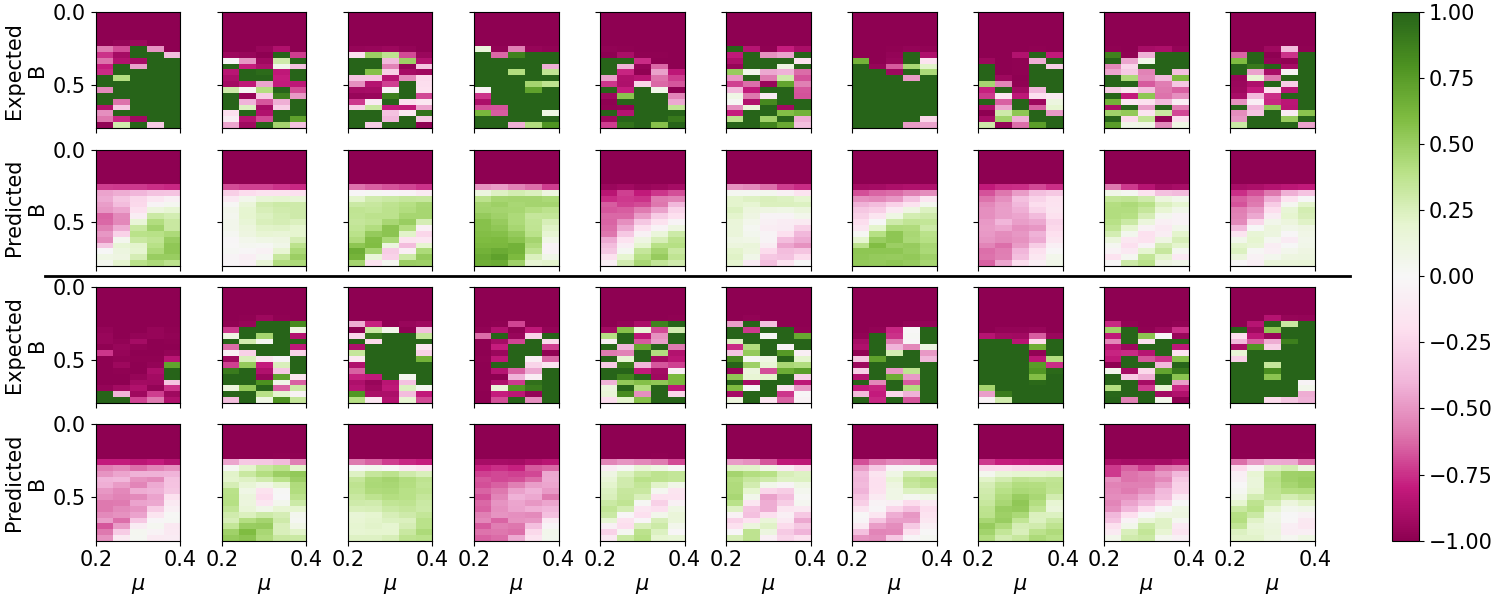}
         \caption{}
     \end{subfigure}
     \caption{Comparison between expected $I_1$ and predicted $I_1^{Pred}$ phase diagrams for the moderate regime. Plots are arranged in vertical pairs where the top plot is the expected $I_1$ and the bottom is the predicted $I_1^{Pred}$. In all plots the phase diagrams are plotted for magnetic field $B$ vs. chemical potential $\mu$ with a continuous $I_1$. Each pair is a different representative sample from the validation set selected at random. }
     \label{fig:I_1ModC}
\end{figure}

\end{minipage}
\clearpage
\twocolumngrid

{\section{Summary and Conclusion}}

In summary, this study presents a comprehensive demonstration of the ability of a Vision Transformer-based neural network to predict Majorana indicators of a 1D nanowire superconductor-semiconductor heterostructure. In fact, the method provides the topological phase diagram even in the challenging (and experimentally relevant) intermediate disorder regime with surprising success. The successful application of this method underscores the growing importance of machine learning in the precise characterization of potential MZMs, which are pivotal for the development of topological quantum computation. These findings not only highlight the potential of these nanowires to serve as building blocks for fault-tolerant quantum computers but also demonstrate that advanced neural networks can significantly enhance the accuracy and efficiency of identifying topological phases, even in the presence of disorder. This work represents a meaningful step towards the practical realization of Majorana-based topological quantum computing, where robust and reliable determination of MZMs is essential. One key feature of our work is a nontrivial generalization of Vision Transformer methodology from a 2D representation to a 3D representation, which is necessitated in our problem by the presence of 3 independent parameters in contrast to 2 in the standard ViT problems.

Current approaches to understanding disorder in these devices heavily rely on heuristic methods, including visually analyzing conductance plots with arbitrary thresholds. These thresholds can be manually adjusted, raising doubts about the reliability of declaring devices as topologically non-trivial. Simulations attempting to validate these heuristics \cite{aghaee2023inas} still leave concerns due to the unknown disorder magnitude and other parameters like the disorder correlation length and spin-orbit coupling. Given that current heuristics essentially involve visual analysis, employing a neural network, particularly Vision Transformers, is a logical next step. These models can process intricate patterns in conductance data, offering a non-arbitrary method to diagnose topological invariants, while leveraging their strength in handling spatial correlations and locality to model the impact of disorder on Majorana nanowires more accurately. We believe that the methodology we develop here can be extended to analyzing many other experimental situations involving quantum computing and condensed matter physics where the data analysis is often hindered by unknown underlying system parameters.  Our work shows that in these situations it may be possible to side-step the unknown parameters by using the vision transformer based machine learning.

Unlike previous heuristics for declaring a device topologically non-trivial, our method provides a way to do so where the probability of an incorrect declaration can not only be known but can be tuned to whatever confidence level is required. By utilizing a continuous prediction approach rather than a categorical classification, our method allows for a more nuanced understanding of the topological phase, enabling researchers to adjust confidence levels based on specific experimental needs. This flexibility is particularly valuable in disordered finite systems, where simple binary classification is typically insufficient. In short disordered wires, merely having a negative topological invariant does not imply the wire can functionally be used for topological computation. By adjusting the topological cutoff, researchers can make the probability of a false declaration of topological non-triviality arbitrarily small.

We consider the topological invariant of a device within a wide variety of disorder regimes, focusing on three in particular: the {extremes regime}, where the disorder is either large or small (most similar to prior results); the full regime, a continuum from low to high disorder; and the most challenging, the moderate disorder regime, where the disorder is fixed at a moderate strength, making it the most difficult to predict but also strongly aligned with current experimental progress. Within each of these regimes, we also examine a wide range of unknown parameters, such as the correlation length and the spin-orbit coupling constant $\alpha$, both of which are very difficult to determine experimentally. Overall, through extensive simulation and analysis, we have demonstrated that our machine learning model can reliably determine the topological nature of a device, even in the presence of significantly varying magnitudes of disorder, correlation lengths, and other unknown model parameters.

In particular, we find that within the {extremes regime}, similar to prior results, even at $C_{cutoff}=0$, we could achieve a probability of $P(T_V < 0|\text{Pass}) = 0.9952$ that a device is topological given it is declared topological. In the full disorder regime, a $C_{cutoff} = 0$ still yields a high $P(T_V < 0|\text{Pass}) = 0.9461$, which can be further increased to $P(T_V < 0|\text{Pass}) = 0.9903$ with a $C_{cutoff} = -0.5$, with only a 17\% reduction in the number of devices declared passing. In the moderate regime, a $C_{cutoff} = 0$ gives $P(T_V < 0|\text{Pass}) = 0.9634$, which can be improved to $P(T_V < 0|\text{Pass}) = 0.9902$ with a $C_{cutoff} = -0.5$, with a 10\% reduction in the percentage of devices that pass. While the neural network in the moderate regime is still effective for declaring devices passing with near-perfect accuracy, achieving $P(T_V < 0|\text{Pass})>0.9998$ with a similar $C_{cutoff} = -0.9$ does come at a cost, as the percentage of devices passing drops 38\%, from 0.9551 to 0.5908. Even targeting $T_V<-0.9$ on a device can be found with $P(T_V < -0.9|\text{Pass})=0.9875$ with a $C_{cutoff}=-0.95$ and a reduction of devices passing dropping only 30\%. 

Going beyond previous methods, we not only demonstrate the ability to predict the aggregate topological invariant of a device, but also provide a method to accurately determine the entire topological phase diagram. This is critically important, as using any 1D nanowire device for its topological properties requires tuning the device to be within the topologically non-trivial regime. In this regard, we achieve errors of $R_{MS}(T_V) = 0.2492$, $R_{MS}(T_V) = 0.2687$, and $R_{MS}(T_V) = 0.4049$ for the extremes, full, and moderate regimes, respectively, showing good accuracy in all cases, both quantitatively and qualitatively.

 Furthermore, the model's performance in the moderate disorder regime, a regime closely aligned with current experimental conditions, demonstrates its practical relevance. In this regime, the model's ability to accurately predict the topological phase with tunable confidence levels suggests that it can be directly applied to ongoing experimental efforts, potentially accelerating the development of robust topological qubits. While the fidelity of the prediction in the moderate regime is lower, it remains highly accurate, especially considering that errors tend to occur at the edges of topological phases, allowing points deep within the phase to be used and trusted. With this in mind, our method, even in its inaccuracies, does so in a useful manner, as the neural network effectively predicts the topological invariant as if it had undergone some Gaussian filtering. This implies that even when the neural network makes incorrect predictions, it often does so in a way that signals uncertainty in the topological invariant, which is significantly superior to simply guessing incorrectly, as might happen in standard classification. The neural network provides a warning that the phase digram may have large errors even while producing a phase diagram. Furthermore, when correct, the method allows for the assessment of the topological invariant within the phase, providing not just the sign but also the magnitude, which is critical in disordered short wires where merely having a negative topological invariant may not be sufficient.

We also consider a first-of-its-kind approach to determining not just the standard topological invariant of a device, but also the ability to determine other Majorana indicators, which are of great importance for determining whether the Majoranas are actually accessible for topological quantum computation. We predict alternative Majorana indicators, such as local density of states (LDOS) based indicators, which are crucial for understanding the spatial distribution of Majoranas along the nanowire. While these indicators proved more challenging to predict, particularly in the presence of complex disorder, the model still provided significant insights. The ability to predict such indicators opens new avenues for assessing the utility of a device for topological quantum computing, beyond merely determining the presence of a topological phase. This is a significant leap, as no practical method for measuring these indicators experimentally, whether by heuristic or otherwise, has been developed besides ours. Since conductance measurements are exceptionally routine and performed on all such devices, our neural network provides the first potential opportunity to utilize these new Majorana indicators in experiments, not just with practical measurements, but with measurements that are already standard in such experiments. In this regard, this work potentially provides a method for bridging the gap from the development of these alternative indicators and demonstration of their utility, to their actual experimental implementation.

In terms of predictions of these LDOS-based Majorana indicators, errors of $R_{MS}(I_2)=0.4640$ and $R_{MS}(I_1)=0.4931$ for the full regime and $R_{MS}(I_2)=0.4799$ and $R_{MS}(I_2)=0.6162$ for the moderate regime were achieved. This shows in both regimes the ability to predict $I_2$ with reasonable accuracy but with a significant reduction in $I_1$ fidelity within the full regime and even more so in the moderate regime. The reason for this is likely the unstable nature of $I_1$ within the moderate regime, combined with the neural network's difficulty with more intricate phase boundaries. In all cases, regardless, qualitative agreement between the phase diagrams remains.

Overall, the results presented in this paper underscore the potential of machine learning, specifically Vision Transformers, to revolutionize the characterization and optimization of topological quantum devices. By providing a method that is both highly accurate and adaptable to a range of disorder conditions, our approach addresses a critical gap in the current toolkit available to experimental physicists for accessing Majorana nanowire devices.
 
 The way we envision this new vision transformer technique could be used to diagnose MZM topology in real experimental nanowires is the following.  Experimental data for the four components of tunnel conductance (the local conductance at each end as well as the two end-to-end nonlocal conductance) should be obtained as a function of the system parameters such as chemical potential (ie gate voltage) and Zeeman field (ie applied magnetic field).  A realistic simulation of this data is then necessary to determine the applicable topological indicators in the training data.  Once the training is done, the machine learning procedure proposed in our work could directly provide the information (with no further simulations) of any new data in new parameter regimes with arbitrary accuracy (as a matter of principle).  It may be particularly useful in the experimental context that our technique could rule out the nontopological parameter regimes so that any eventual braiding measurements focus on the suitable parameter regime.  Since the total parameter regime (gate voltage, magnetic field, wire length...) is huge, any information on narrowing the potential topological regime for experimental samples should be a huge help in experimental progress.
 
In future work, we expect it may be possible to improve our method by incorporating additional data, such as data resulting from varying barrier voltages used within the topological gap protocol, which might provide extra information for the ML scheme, enhancing the accuracy of Majorana indicator predictions. Scaling up may also yield better results with more measurements in $B$ and $\mu$. Further, interestingly while not shown here we found that our results could be significantly improved by more precisely narrowing the range of the disorder correlation length, so in future work if this could be done, a more narrowly trained neural network will yield much better results. It may also be possible to tune wires to optimize for these indicators by using our ML scheme, though this would likely require significant additional training data related to predicting indicators for potentials modified by tuning gates. We believe that our work paves the way for Vision Transformers to become an integral part of topological data analysis facilitating the prediction of topological versus trivial parameter regimes in experimental samples.

\section{Acknowledgement}
This work was supported by the Laboratory for Physical Sciences.  We thank Jay Deep Sau for many helpful discussions. We also thank UMD HPC Zaratan for computational resources provided.

\bibliography{mainbib}
\end{document}